\documentclass[a4paper,11pt]{article}
\usepackage[utf8]{inputenc}

\usepackage{jcappub} % for details on the use of the package, please see the JINST-author-manual

\usepackage{lineno}
\usepackage{orcidlink}
\usepackage{booktabs}
\usepackage{float}
\usepackage{subfig}
\usepackage{graphicx}% Include figure files
\usepackage{dcolumn}% Align table columns on decimal point
\usepackage{rotating}
\usepackage{bm}% bold math
\usepackage{hyperref}% add hypertext capabilities
\arxivnumber{2503.17206} % Only if you have one
\title{\boldmath Dark Matter (S)pins the Planet}

\author[a,b]{Haihao Shi\orcidlink{0009-0007-9418-2632}}\note{Haihao Shi and Junda Zhou contribute equally to this work;}

 \emailAdd{shihaihao@xao.ac.cn}
\affiliation[a]{Xinjiang Astronomical Observatory, Chinese Academy of Sciences, Urumqi 830011, People's Republic of China}

\affiliation[b]{College of Astronomy and Space Science, University of Chinese Academy of Sciences, Beijing 101408, People's Republic of China}
\author[c,b]{Junda Zhou\orcidlink{0009-0004-7473-1727}}
 \emailAdd{zhoujunda@ynao.ac.cn}
\affiliation[c]{Yunnan Observatories, Chinese Academy of Sciences,  Kunming, 650216, People's Republic of China}
\author[a,b]{Zhenyang Huang\orcidlink{0009-0009-8188-5632}}%
 \emailAdd{huangzhenyang@xao.ac.cn}
\author[d,a]{Guoliang L\"u\orcidlink{0000-0002-3839-4864} }
\note{Corresponding author:Guoliang L\"u(guolianglv@xao.ac.cn);}
\emailAdd{guolianglv@xao.ac.cn}

\affiliation[d]{School of Physical Science and Technology, Xinjiang University, Urumqi 830046, People's Republic of China}
\author[c,b,e,f]{Xuefei Chen \orcidlink{0000-0001-5284-8001}}%
\note{Corresponding author:Xuefei Chen (cxf@ynao.ac.cn)}
 \emailAdd{cxf@ynao.ac.cn}
\affiliation[e]{Key Laboratory for Structure and Evolution of Celestial Objects, Chinese Academy of Sciences, Kunming 650216, People's Republic of China}

\affiliation[f]{International Centre of Supernovae, Yunnan Key Laboratory, Kunming 650216, People's Republic of China}

\abstract{Dark matter heating in planets has been proposed as a potential probe for dark matter detection. Assuming near-equilibrium conditions, we find that the energy input from dark matter raises planetary temperatures and accelerates rotation. The distribution of energy between heating and rotational acceleration depends on both planetary properties and external inputs, suggesting that previous studies may have overestimated the heating contribution. At high dark matter densities, planetary rotation stabilizes earlier and becomes primarily governed by dark matter effects.}

\begin{document}
\maketitle
\flushbottom

\section{Introduction}

Dark matter (DM) constitutes approximately 85\% of the total matter in the Universe \citep{Planck:2018vyg}, as supported by a multitude of astrophysical and cosmological observations \citep{WMAP:2012nax,Bergstrom:2000pn,Young:2016ala,Billard:2021uyg}. Despite its pervasive presence, the fundamental nature and composition of dark matter particles remain elusive, pointing to physics beyond the Standard Model and general relativity. Numerous candidates have been proposed, including axions \citep{Wilczek:1977pj,PRESKILL1983127,1983PhLB..120..137D,Abbott:1982af,Svrcek:2006yi}, fuzzy dark matter \citep{Hu:2000ke,Hui:2021tkt,Burkert:2020laq}, primordial black holes \citep{Lacki_2010,Carr:1974nx,Carr:2016drx}, and so on. Detection strategies span a wide range of scales—from microscopic accelerator-based experiments to constraints derived from cosmic-scale astronomical observations \cite{PhysRevD.70.103517,Bertone:2004pz,Safronova:2017xyt,Alekhin:2015byh,ADMX:2009iij,Goodman:2010ku,Fermi-LAT:2016uux,Nesti:2023tid,Bertone:2018krk}. Recently, increasing attention has been paid to constraining dark matter using planetary-scale systems, including Earth \cite{PhysRevD.76.043523,PhysRevD.94.035024,PhysRevD.101.043001}, Jupiter \citep{10.1143/ptp/87.3.685,Blanco:2023qgi,Blanco:2023qgi}, white dwarves \citep{PhysRevD.81.083520,PhysRevD.98.115027,PhysRevD.100.043020,Krall_2018,Dasgupta_2019}, neutron stars \citep{PhysRevD.40.3221,Gould:1989gw,PhysRevD.77.023006,PhysRevD.77.043515,PhysRevD.107.115016,PhysRevD.81.123521,PhysRevD.82.063531,PhysRevD.85.023519,PhysRevLett.107.091301,Guver_2014,PhysRevD.87.055012,PhysRevD.87.123507,PhysRevD.88.123505,PhysRevD.83.083512,PhysRevLett.115.141301,PhysRevD.92.063007,PhysRevD.94.063001},exoplanets \cite{Leane:2020wob,Benito:2024yki,Stephens:2021sfq,Phoroutan-Mehr:2025hjz} and so on. Studying dark matter interactions on planetary scales is particularly compelling, as it provides a unique bridge between astrophysical observations and particle physics experiments \citep{Smith:2019wgb}. Planets, having coexisted with the galactic dark matter halo for billions of years, serve as long-term integrators of dark matter effects. These interactions may lead to cumulative and potentially observable consequences, such as changes in planetary temperature, rotational dynamics, and atmospheric properties. Furthermore, dark matter effects at planetary scales could influence planetary habitability by altering thermal conditions, potentially affecting the stability of liquid water and atmospheric evolution \citep{Hooper:2011dw}. With upcoming exoplanet surveys, such as James Webb Space Telescope \citep{Gardner:2006ky,Baryakhtar:2017dbj} and Transiting Exoplanet Survey Satellite \citep{Deming:2009is,2024arXiv241012905W}, providing increasingly precise planetary data, the study of dark matter at planetary scales will become a crucial aspect of both dark matter detection and habitability assessment.

The study of planetary dark matter capture represents a key approach to probing dark matter at planetary scales. While the capture process has been widely examined, previous research has primarily focused on its thermal effects, particularly on planetary temperature \citep{Adler:2008ky,Bramante:2022pmn,Chauhan:2016joa}. However, the transition from the thermodynamic effects of dark matter to its influence on planetary dynamics warrants attention. Landau once discussed the influence of thermodynamic properties, such as entropy, on the macroscopic dynamics of a system \citep{landau2013statistical}. Applying similar concepts to the interaction between dark matter and planets is both natural and significant. Planets possess distinct characteristics: they move through cosmic space, rotate, and gradually evolve toward near-thermodynamic equilibrium over long timescales. Both thermodynamic and dynamical properties—such as temperature and rotation rate—play critical roles in planetary evolution, influencing atmospheric composition and surface morphology, which are central to assessments of planetary habitability.

This paper begins with a brief overview of the dark matter heating mechanism in \autoref{Energy Injection Mechanism}. In \autoref{Rotation Acceleration}, we extend dark matter heating to its effect on planetary rotational angular velocity. In \autoref{simulation}, we conduct evolutionary simulations for exoplanets, including Epsilon Eridani b, listed in \autoref{exoplanets list}. In \autoref{obs}, we shift our focus to Jupiter and Earth, providing predictions of dark matter’s potential impact on these planets. Finally, \autoref{conclusion} presents a summary and outlook.

\section{Dark Matter Heating}\label{Energy Injection Mechanism}

\textcolor{black}{We focus here on the thermal effects of dark matter heating, a process fundamentally distinct from mechanical effects like kinetic energy transfer. This heating arises from the scattering, capture, and subsequent annihilation of dark matter particles within an exoplanet's core. This process converts the particles' rest-mass energy into a localized heat source that is then absorbed by the planetary interior. }Following the framework established in \cite{PhysRevLett.126.161101}, we assume equilibrium between dark matter scattering and annihilation processes. The resulting heat flux depends on the fraction of incident dark matter particles captured from the external flux reservoir. For the dark matter energy injection power, it is equal to the dark matter heat power mentioned in \cite{PhysRevLett.126.161101}:

\begin{equation}\label{heatDM}  
\Gamma_{\text{heat}}^{\text{DM}} = f \pi R^2 \rho_\chi(r) v_0 \left( 1 + \frac{3 v_{\text{esc}}^2}{2 v_d(r)^2} \right).  
\end{equation}

Here, $f$ is the fraction of captured DM particles that have passed through, $R$ is the planetary radius, and $\rho_\chi(r)$ represents the dark matter density. The average speed in the DM rest frame, \( v_0 \), is related to the velocity dispersion \( v_d(r) \) as \( v_0 = \sqrt{\frac{8}{3\pi}} v_d(r) \) at a distance \( r \) from the Galactic Center. The escape velocity is given by \( v_{\text{esc}}^2 = \frac{2G M}{R} \). The circular velocity \( v_c(r) \) in the galaxy is related to the DM velocity dispersion by \( v_d(r) = \sqrt{\frac{3}{2}} v_c(r) \). We extract the circular velocities at different radii in the Milky Way by combining data from the gas, bulge, and disk components, along with the analytic expressions for DM contributions to the total velocity from \cite{10.1093/mnras/stz1698}.

%For the Solar System neighborhood, given the parameters:  
%\(\rho_\chi(r_{\odot}) = 0.38 \,\text{GeV/cm}^{3} \sim 6.76 \times 10^{-28} \,\text{kg/cm}^{3}\),  
%adopting the NFW model with \(v_h = 430\,\text{km/s}\) and \(r_0 = 9.4\,\text{kpc}\), we obtain  
%\( v_{d(\odot)} = v_h \sqrt{\frac{r_0}{r_{\odot}} \ln \left( \frac{r_{\odot} + r_0}{r_0} \right) - \frac{r_0}{r_{\odot} + r_0}} \sim 185\,\text{km/s} \),  
%\( v_{0} = 170.4\,\text{km/s} \), \( v_{c(r_{\odot})} = 151\,\text{km/s} \), and \( v_{\text{esc}}^{2} = 3600\,\text{km}^{2}/\text{s}^{2} \).  

\textcolor{black}{Next, we calculate the rate of energy supplied to the planet by dark matter, as described in \autoref{heatDM}. This quantity is determined from the capture and subsequent interactions of dark matter particles within the planetary interior, which can deposit energy through scattering, annihilation, or decay processes. In the following discussion, we denote \(\Gamma_{\text{heat}}^{\text{DM}}\) as \(\Gamma_{\text{in}}^{\text{DM}}\) to emphasize that the incoming energy from dark matter is not necessarily fully converted into thermal energy. Part of this energy may be redistributed into other channels, which in the present study are manifested as changes in the planet’s rotational state, a central focus of our analysis. This distinction is essential for correctly relating the thermal effects of dark matter to its potential impact on planetary dynamics under near-equilibrium conditions.}

\section{From Dark Matter Heating to Dark Matter Spin-up}\label{Rotation Acceleration}

\textcolor{black}{The dynamical state of a macroscopic body in thermodynamic equilibrium, as discussed in the ``Macroscopic motion\footnote{\url{https://www.sciencedirect.com/science/article/pii/B9780080570464500099}}'' section of \cite{landau2013statistical}, is restricted to uniform translational and rotational motion.} The corresponding translational and angular velocities are given by: 

\begin{equation}\label{eqa}
    \begin{split}
        u &= aT ,\\
        \Omega &= bT,
    \end{split}
\end{equation}
respectively, where \( a \) and \( b \) are dimensional constants dependent on the system's state, and \( T \) is the equilibrium temperature.

% This conclusion applies to certain planetary systems that have evolved long enough to transition from an initial non-equilibrium state to a near-equilibrium state (NES) \citep{Moreau1990} and are progressing toward a Non-equilibrium Steady State \citep{BIANCA2012359} . Since the energy input from dark matter and the host star remains stable over long timescales, it does not significantly perturb the planet's NES. This can be equivalently described as an additional temperature term, beyond the cosmic background temperature, evolves slowly with the environment. Consequently, the planetary system's approach to thermodynamic equilibrium naturally follows this evolution. Under this assumption, the conclusions from \cite{landau2013statistical} remain applicable.

\textcolor{black}{Our conclusions apply to mature planetary systems that have evolved long enough to reach a quasi-static equilibrium. This state is maintained because the primary energy inputs—from the host star and ambient dark matter—are stable over the long timescales of planetary evolution, and therefore do not significantly perturb the system.} \textcolor{black}{Under these conditions, the planet's evolution can be viewed as passing through a continuous sequence of thermodynamic equilibrium states. This justifies the application of standard statistical mechanics principles, such as those described in \cite{landau2013statistical}, to model its behavior.}

Here, we do not consider translational motion, allowing us to focus on the effect of increased temperature on rotational motion. This approach provides an upper bound on the spin acceleration without being constrained by the ratio of translational to rotational velocity, as this ratio can clearly be treated as a free parameter. In this idealized model, incorporating the dark matter energy injection mechanism introduced in the previous section is the central topic of this work.

\begin{figure}[H]
    \centering
    \includegraphics[width=0.8\textwidth]{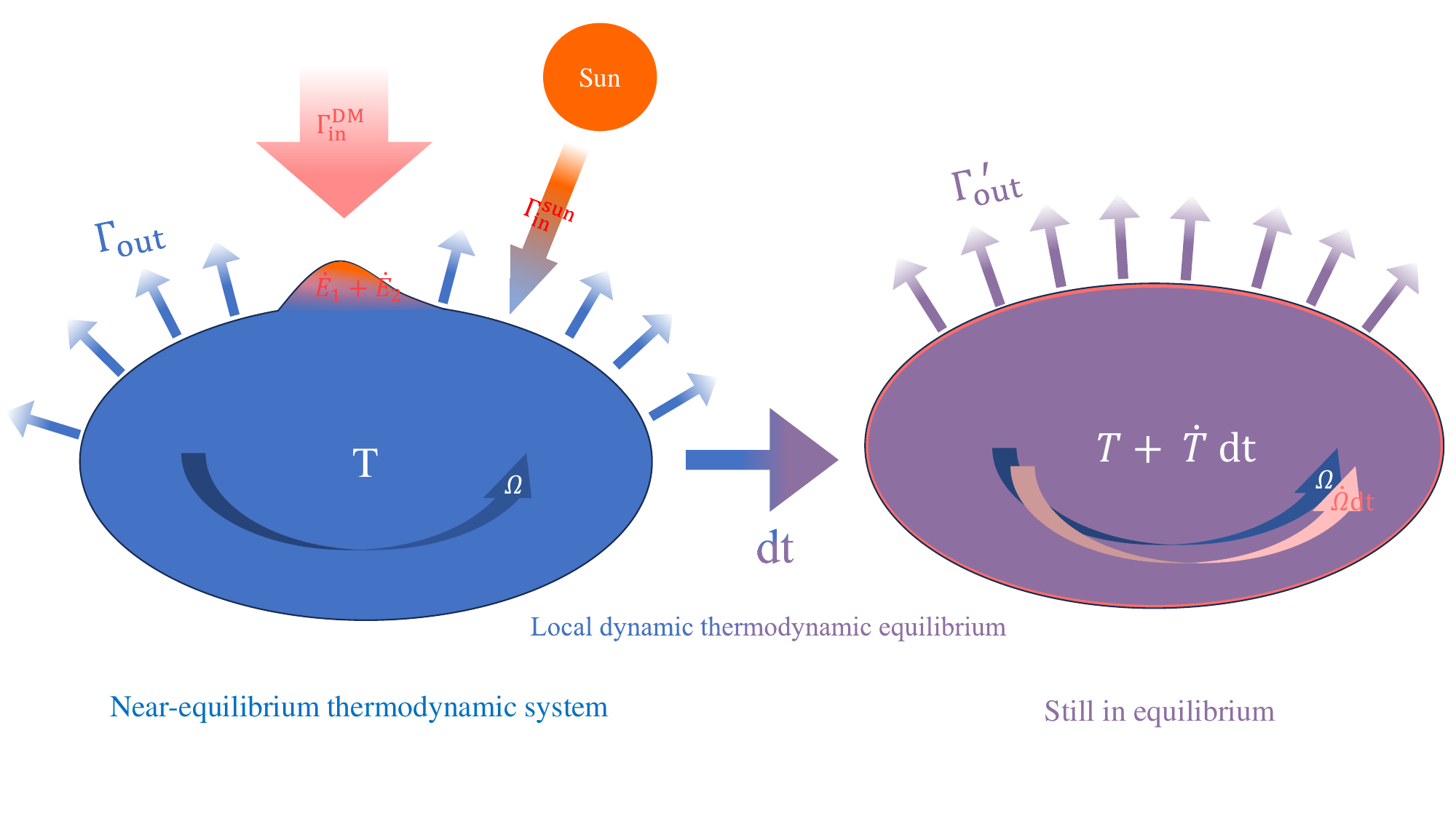}  % Replace with the actual image name and extension
    \caption{The left and right figures respectively represent the planetary system before and after experiencing perturbation $\dot T$ due to dark matter and sun's heating. Since the perturbation is minimal, we consider the process under local dynamic thermodynamic equilibrium. The perturbations follow \(\dot E_{1} + \dot E_{2}  \rightarrow \dot T + \dot \Omega\), where the planetary temperature evolves as \(T \rightarrow T + \dot T\mathrm{d}t\), and the planetary angular velocity changes as \(\Omega\rightarrow \Omega + \dot \Omega\mathrm{d}t\).}
    \label{spin model}
\end{figure}

% The structure of our model is illustrated in \autoref{spin model}, where the power inputs from dark matter and the host star, \(\Gamma_{\text{in}}^{\text{DM}}\) and \(\Gamma_{\text{in}}^{\text{sun}}\), are converted into temperature and angular velocity changes, \(\dot{T}\) and \(\dot{\Omega}\), respectively. In this process, the planetary temperature evolves from \(T_{i}\) to \(T_{i+1}=T_{i} + \dot{T_{i}} \mathrm{d}t\), while the angular velocity changes from \(\Omega\) to \(\Omega_{i+1}=\Omega_{i} + \dot{\Omega_{i}} \mathrm{d}t\), with the relation.

\textcolor{black}{The structure of our model is schematically shown in \autoref{spin model}. In this framework, the energy input rates from dark matter and the host star, \(\Gamma_{\text{in}}^{\text{DM}}\) and \(\Gamma_{\text{in}}^{\text{sun}}\), act as external power sources that drive the evolution of the planet’s thermodynamic and dynamical states. Specifically, these inputs are converted into changes in the planetary temperature and angular velocity, \(\dot{T}\) and \(\dot{\Omega}\), respectively. Over each integration step \(\mathrm{d}t\), the temperature evolves from \(T_{i}\) to
\begin{equation}\label{Ti}
T_{i+1} = T_{i} + \dot{T}_{i}\,\mathrm{d}t ,
\end{equation}
while the angular velocity evolves from \(\Omega_{i}\) to
\begin{equation}\label{Oi}
\Omega_{i+1} = \Omega_{i} + \dot{\Omega}_{i}\,\mathrm{d}t .
\end{equation}
The quantities \(\dot{T}\) and \(\dot{\Omega}\) are related via \autoref{eqb}, which is derived from \autoref{eqa} and couples the thermal and rotational evolution in our near-equilibrium planetary model.}

\textcolor{black}{
\begin{equation}\label{eqb}
    \dot{\Omega_{i}} = b_{i} \dot{T_{i}}.
\end{equation}
}

\textcolor{black}{Here, \text{i} represents the i-th state. }Over a time interval \(\mathrm{d}t\), the power provided by dark matter and sun is \(\Gamma_\text{in} = \Gamma_\text{in}^\text{DM} + \Gamma_\text{in}^\text{sun}\), which is partitioned into two components: \(\dot{E}_1\), contributing to internal energy (heating), and \(\dot{E}_2\), increasing the angular velocity.  \textcolor{black}{In our numerical model, the parameter \(b\) is updated at each time step. 
At step \(i\), \(b_i\) is computed from the initial state \((\Omega_i, T_i)\) via  
\(b_i = \Omega_i / T_i\), and is then treated as constant during that step, serving as a key parameter governing the evolution of temperature and angular velocity (see \autoref{Ti} $\sim$ \autoref{eqb}).  
Once the next-step state \((\Omega_{i+1}, T_{i+1})\) is obtained, \(b_{i+1}\) is recalculated for the following step.  
Under the near-equilibrium hypothesis, however, \(b\) should remain invariant throughout the evolution, independent of time. Indeed,  
\[
b_{i+1}=
\frac{\Omega_{i+1}}{T_{i+1}}
= \frac{\Omega_{i} + \dot{\Omega}_{i} \,\mathrm{d}t}{T_{i} + \dot{T}_{i} \,\mathrm{d}t} 
= \frac{b_{i}\,(T_{i} + \dot{T}_{i} \,\mathrm{d}t)}{T_{i} + \dot{T}_{i} \,\mathrm{d}t} 
= b_{i},
\]
showing that \(b\) remains constant during the evolution, consistent with Landau’s conclusion.  
This constancy is also confirmed in our detailed numerical simulations, thereby validating the self-consistency and physical soundness of the near-equilibrium assumption.  
To emphasize this property, we will hereafter denote it simply as \(b\) without a time-step index.}

\textcolor{black}{Since the planet remains in dynamic radiative equilibrium during long-timescale evolution}, the energy balance equation is given by

\begin{equation}\label{eqc}
    \Gamma_\text{in} = \Gamma_\text{out} + I\Omega\dot\Omega,
\end{equation}
\textcolor{black}{Here, \(\Gamma_{\text{in}}\) denotes the total incoming energy from all sources, which in this case is 
\begin{equation}
\Gamma_\text{in} = \Gamma_\text{in}^\text{DM} + \Gamma_\text{in}^\text{sun} ,
\end{equation}
where \(\Gamma_\text{in}^\text{DM}\) and \(\Gamma_\text{in}^\text{sun}\) are the contributions from dark matter and the host star, respectively. The term \(\Gamma_{\text{out}}\) represents the total energy loss, modeled under the single-layer atmospheric approximation as planetary blackbody radiation, 
\begin{equation}
\Gamma_{\text{out}} = 4\pi R^2\sigma T'^4 ,
\end{equation}
with \(T'\) being the planetary temperature after a time step \(\mathrm{d}t\), given by \(T' = T_{i+1}\) according to \autoref{Ti}. The quantity \(I\) denotes the planetary moment of inertia.}

Solving equations \autoref{eqa}, \autoref{eqb}, and \autoref{eqc} simultaneously, with the assumption: \(T\gg \dot T\), we obtain

\begin{equation}  
\begin{aligned}
\dot T  &= \frac{\Gamma_\text{in}-4\pi R^2\sigma T^4}{16\pi R^2\sigma T^3 \mathrm{d}t+I\Omega b} ,\\
\dot \Omega &= b\dot T.
\end{aligned}
\label{res1}
\end{equation}

Here, \(\mathrm{d}t\) in \autoref{res1} is recorded to assess the adequacy of the chosen time step \(\Delta t\). 
If \(\Delta t\) is sufficiently small such that further reductions lead to negligible differences in the simulation results, this confirms that the temporal resolution is properly resolved and the chosen \(\Delta t\) is appropriate. When \(\Gamma_\text{in} > \Gamma_\text{out}\), both \(\dot{E}_1\) and \(\dot{E}_2\) exist and are given by

\begin{equation}  
\begin{aligned}
\dot E_1 &= \Gamma_\text{in} - \dot E_2,\\
\dot E_2 &= I\Omega b \frac{\Gamma_\text{in}-4\pi R^2\sigma T^4}{16\pi R^2\sigma T^3 \mathrm{d}t+I\Omega b}.
\end{aligned}
\label{res2}
\end{equation}

\(\dot{E}_{2}\) represents the portion of the external power input that contributes to the change in the planet's angular velocity. It can be observed that when \(\Gamma_\text{in} = 4\pi R^2\sigma T^4\), the system reaches a Non-equilibrium Steady State, meaning that the statistical physical parameters describing the system on macroscopic scales remain constant over time, despite the system not being in thermal equilibrium. Notably, dynamical properties such as the rotational angular velocity also remain unchanged in this state.

\textcolor{black}{The partitioning of injected dark matter (DM) energy into thermal (heating) and rotational (spinning) components is not a pre-defined assumption in our model. Instead, this energy branching ratio emerges self-consistently from the solution of the coupled evolution equations (\autoref{eqa}$\sim$\autoref{eqc}), directly reflecting the planet's thermodynamic and dynamical response under the assumption of dynamic radiative equilibrium.} Specifically, we solve the full evolution equations with DM energy as a source term. The resulting division between heating and spin-up is an outcome of the integrated evolution and depends on planetary properties at each stage. This dynamical regime—i.e., the balance between internal and rotational energy gain—is not prescribed but determined by the model itself. Our analysis of energy distribution is based on these evolved solutions.

\textcolor{black}{To provide an intuitive physical picture, we begin with the first law of thermodynamics for a rotating body in the quasi-adiabatic limit. Neglecting $P\,\mathrm dV$ work because we are not considering volume changes, it reads
\[
\mathrm dE \;=\; T\,\mathrm dS \;+\; \Omega\,\mathrm dJ .
\]
Here, \(E\) is the energy, \(S\) its entropy, and \(J\) its angular momentum; 
\(T\) is the equilibrium temperature (thermodynamic conjugate to \(S\)), \(\Omega\) the bulk angular velocity
(conjugate to \(J\)).}

\textcolor{black}{Moreover, using the equilibrium relation between temperature and angular velocity from Ref.~\cite{landau2013statistical}, $\Omega = b\,T$, the two variables are thermodynamically coupled through the parameter $b$. Hence, when DM heating supplies energy, the increment $\mathrm dE$ is partitioned between an entropic/thermal part $T\,\mathrm dS$ and a rotational part $\Omega\,\mathrm dJ$ in proportions governed by $b$ along the sequence of near-equilibrium states. }

\textcolor{black}{Importantly, from a dynamical standpoint, energy transfer always requires a coupling medium. In our context, the conversion of injected energy into the thermal ($T,\mathrm dS$) and rotational ($\Omega,\mathrm dJ$) parts necessarily proceeds through physical channels—here we adopt the simplest but effective picture of a "heat flow", while acknowledging that its concrete realization can be complex and system-dependent. Their relative efficiencies are likewise system-dependent and model-specific, but the ultimate asymptotic effects on planetary evolution timescales still tend to be as described in Ref.~\cite{landau2013statistical}. In the present work, we rely only on the assumption that such a weak coupling exists and drives a quasi-adiabatic sequence, while the microphysical details of the coupling are beyond our scope. While we do not investigate the microphysical details of this coupling, this does not affect our main idea: potential heating from dark matter does not convert all its energy into an increase in the system's temperature. }

\section{Simulation}\label{simulation}

Based on the theory proposed in \autoref{Rotation Acceleration}, we simulate the 15 planets listed in \autoref{tab:exoplanets} and analyze their temperature evolution under \(\rho_{\text{local}} = 0.38 \text{GeV/cm}^{3}\) \citep{deSalas_2021} and \(10^{4} \rho_{\text{local}}\). For the treatment of \(\Gamma_\text{in}^\text{sun}\), given the mass \(M_s\) and age \(t\) of the host star (or the Sun), we interpolate the data from \citep{2015AA...577A..42B} to obtain the stellar luminosity \(L_s\). Assuming no planetary reflection, the stellar energy input is given by \(\Gamma_{\text{in}}^{\text{sun}} = L_s R^2 / 4D^2\), where \(R\) is the planetary radius and \(D\) is the orbital distance, assuming a circular orbit. Since the rotational velocity of exoplanets is not directly observable, this section does not consider specific changes in rotation rate but instead focuses on the ratio of energy used for rotational changes to the total energy input, given by \(\frac{\dot{E_{2}}}{\dot{E_{1}}+\dot{E_{2}}}\).

% \onecolumngrid

\begin{figure}[H]
    \centering
    % First subplot
    \subfloat[]{
        \includegraphics[scale=0.38]{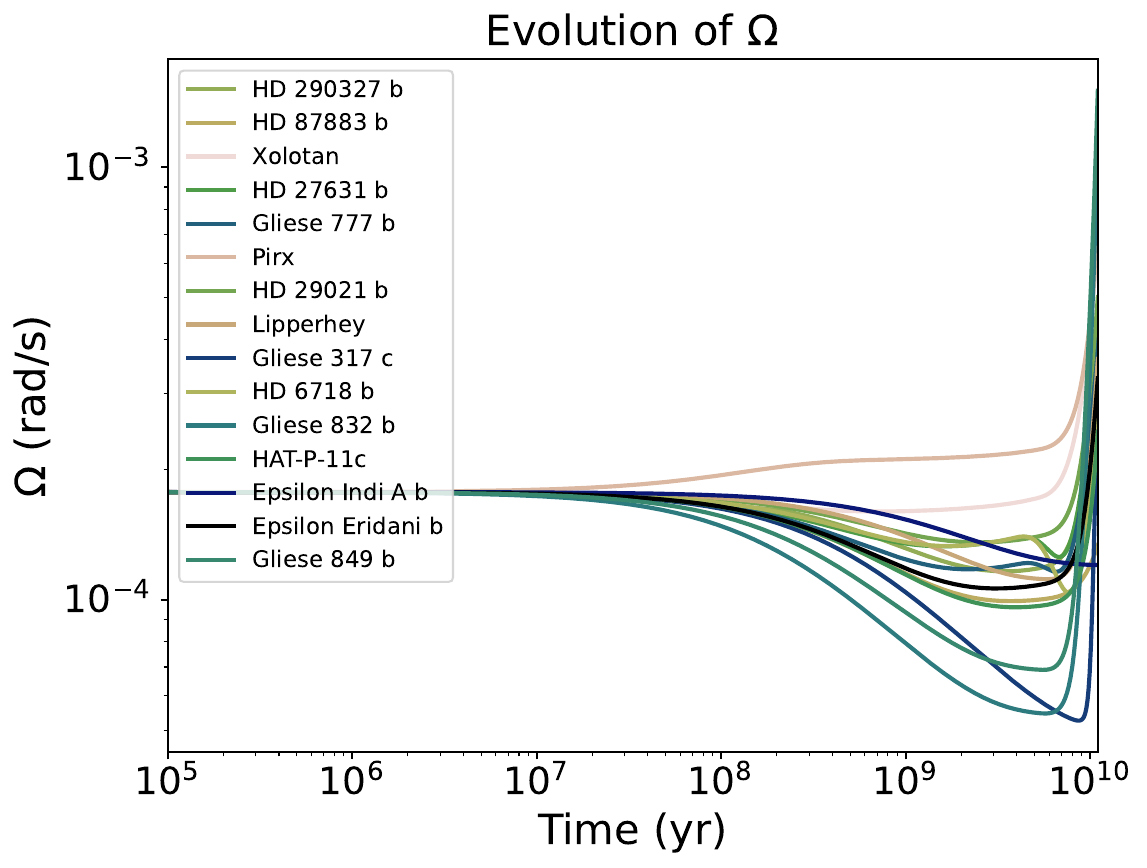}
        \label{sa}
    }
    % Second subplot
    \subfloat[]{
        \includegraphics[scale=0.38]{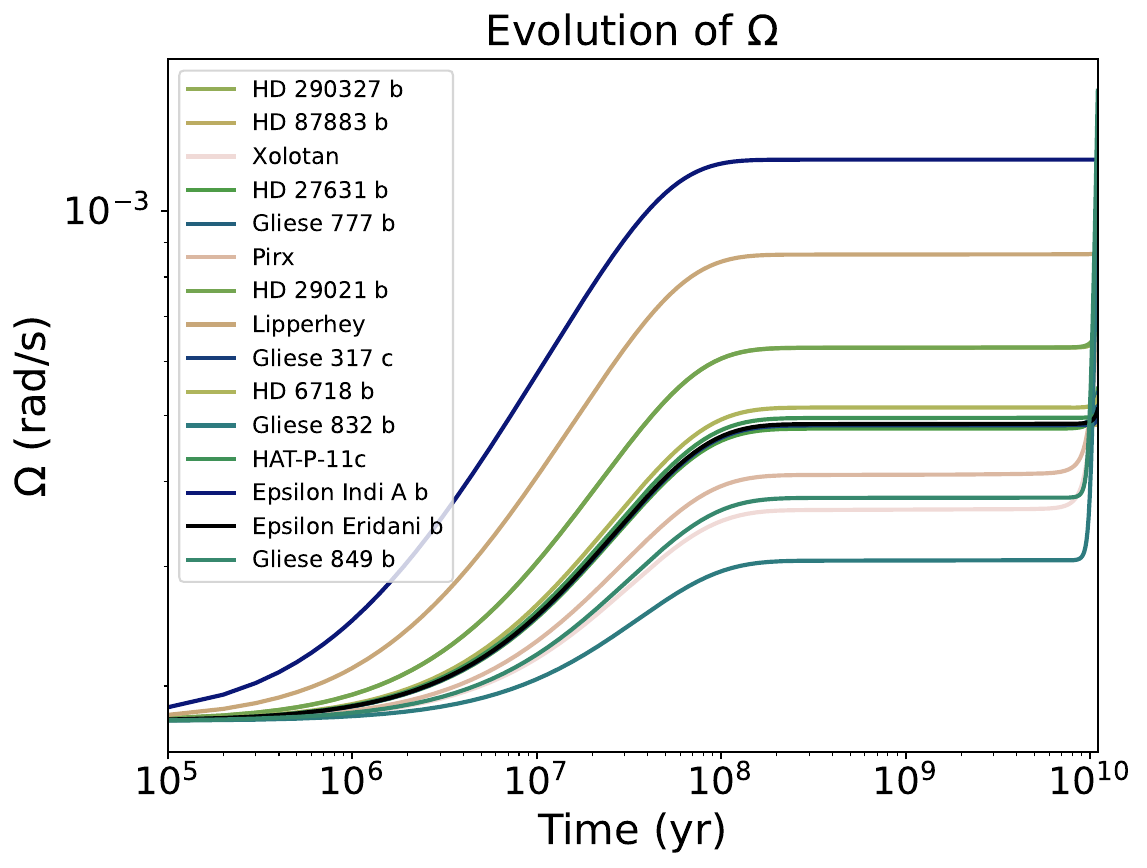}
        \label{sb}
    }
    \caption{Angular velocity evolution of the planets under different dark matter densities. \autoref{sa} shows the evolution at \(\rho_{\text{local}}\), while \autoref{sb} corresponds to \(10^{4} \rho_{\text{local}}\). The time origin here is set at the current age of the star. We set the initial angular velocity of each simulated planet to that of Jupiter.}
    \label{15omega}
\end{figure}

It can be observed that in low dark matter density environments \textcolor{black}{(The DM density in \autoref{sa} is set to \(\rho_{\text{local}}\))}, the planetary angular velocity is largely determined by stellar evolution. In high dark matter density environments \textcolor{black}{(The DM density in \autoref{sb} is set to \(10^{4}\times \rho_{\text{local}}\))}, where \(\Gamma_\text{in}^\text{DM} \gg \Gamma_\text{in}^\text{sun}\), the angular velocity stabilizes at a constant value upon reaching the Non-equilibrium Steady State, which is the origin of the term "pins" in our title. It should be noted that the sudden change in \autoref{15omega} after \(10^{10} \text{ yr}\) is due to the termination of the stellar model at this point, which depends on our understanding of stellar evolution theory.

Our results indicate that the heating effect of dark matter on planetary temperature depends on both the dark matter density and the intrinsic properties of the planet. Moreover, the resulting change in rotational velocity due to the shift in effective temperature at equilibrium reduces the overall impact, making it less significant than suggested in previous studies \citep{Leane:2020wob,2024arXiv240502393A,Phoroutan-Mehr:2025hjz}. Further simulation results are shown in \autoref{simulation_result}.
It should be noted that this analysis is based on a simplified toy model. In reality, factors such as planetary obliquity and internal heat transport would introduce additional complexity. However, the aim of this work is not to provide precise predictions, but rather to demonstrate that the heating effect of dark matter may be mitigated by the rotational acceleration mechanism proposed in this study.

\section{Solar System}\label{obs}

Let us now turn our attention back to the Solar System. Since our understanding of Solar System planets, such as Jupiter and Earth, is far more comprehensive than that of exoplanets, our model allows for more detailed predictions that may be tested in the future. Due to space limitations, the temperature and angular velocity evolution of Jupiter and Earth are presented in \autoref{solar_simu}.  

We find that, compared to Earth, the effect of our mechanism on Jupiter is less significant due to its larger mass and volume. However, for Earth, our model predicts that the combined energy input from dark matter and solar radiation would raise the effective temperature by an amount on the order of $10^{-2} \mathrm{~K}$ over 100 years, and $10^{-1} \mathrm{~K}$ over 1000 years. Additionally, the Earth's rotational velocity is expected to increase by about \(10^{-8} \text{ rad/s}\) over 100 years and \(10^{-7} \text{ rad/s}\) over 1000 years. This implies that, under the proposed mechanism, Earth’s day length would shorten by approximately 12 seconds in 100 years and by 120 seconds in 1000 years, assuming a current rotation period of 24 hours. For Earth, we predict that the heating of dark matter will accelerate its rotation period on the order of seconds per hundred years. This may be observable by ground-based measurement methods. However, the Earth's rotational speed is also affected by a series of other effects, such as tidal effects, earthquakes, and so on. Separating the heating effect of dark matter is difficult. It is important to emphasize that this study focuses on changes in the effective temperature, rather than the temperature of the human living environment or any specific atmospheric layer. Any attempt to associate this with phenomena such as the greenhouse effect would be inappropriate. For Jupiter, our model did not use all the energy provided by dark matter for heating, but instead allocated a portion to the planet's rotational energy. Therefore, in our calculations based on radiation balance, Jupiter's temperature will gradually decrease. This is consistent with the actual situation. The current temperature of Jupiter is not entirely maintained by solar radiation, but includes nuclear fission reactions inside Jupiter, which have not been considered by our model.

Additionally, we explore the impact of different fractions of captured dark matter (\(f\)) on planets, as this may provide constraints on dark matter parameters at earth. Unfortunately, as shown in \autoref{fig:earth_f_sun+DM} and \autoref{fig:earth_f_onlyDM}, the influence of \(f\) on planets within the Solar System is not significant. However, future advancements in our understanding of exoplanets may offer more information for constraining dark matter parameters.

% \textcolor{black}{\section{Discussion}}

\section{Conclusion}\label{conclusion}

In this work, based on \cite{landau2013statistical} and the near-equilibrium state assumption, we derive the time evolution of planetary temperature and angular velocity under the combined energy input from dark matter and the host star. We apply our model to Jupiter, Earth, and the exoplanets listed in \autoref{exoplanets list}. Our theory suggests that the energy provided by dark matter heating is not entirely converted into temperature but is distributed according to the planet's intrinsic properties, such as mass and radius, as well as its current state, including temperature and angular velocity. Importantly, this indicates that the effect of dark matter heating has been overestimated in previous studies.

Based on our model calculations, we predict that Earth's effective temperature will increase by an amount on the order of $10^{-2} \mathrm{~K}$ over 100 years and $10^{-1} \mathrm{~K}$ over 1000 years. It should be reemphasized that this refers to the effective temperature derived from radiative energy balance, not the temperature of the human environment or any specific atmospheric layer. Accordingly, this effect should not be conflated with the greenhouse effect or global warming. The rotational velocity is expected to increase by about \(10^{-8} \text{ rad/s}\) over 100 years and \(10^{-7} \text{ rad/s}\) over 1000 years. This implies that Earth's rotation period will shorten by approximately 12 seconds over 100 years and by 120 seconds over 1000 years due to the proposed mechanism. In regions with higher dark matter densities, such as closer to the Galactic center, planetary angular velocities may be "spun" by dark matter and then "pinned" at a fixed value. Finally, we analyze the impact of different fractions of captured dark matter (\(f\)) on Earth and find that variations in \(f\) are effectively indistinguishable. At present, we cannot use data from Earth are insufficient to constrain the dark matter \(f\) parameter.

In the future, as the quest for a second home among the stars unfolds, the dark matter-induced rotational effects explored in this work may provide a useful reference for evaluating planetary habitability.
\begin{acknowledgments}

I would like to thank Qiyu Yan and  Xuwei Zhang for the useful discussions during the research process. This work received support from the National Natural Science Foundation of China under grants 12288102, 12373038, 12125303, 12090040/3, and U2031204; the Natural Science Foundation of Xinjiang No. 2022TSYCLJ0006; the science research grants from the China Manned Space Project No. CMS-CSST-2021-A10; the National Key R\&D Program of China No. 2021YFA1600401 and No. 2021YFA1600403; the Natural Science Foundation of Yunnan Province Nos. 202201BC070003 and 202001AW070007; the International Centre of Supernovae, Yunnan Key Laboratory No. 202302AN360001; and the Yunnan Revitalization Talent Support Program $-$ Science \& Technology Champion Project No. 202305AB350003.

\end{acknowledgments}


\bibliographystyle{apsrev4-2} % 按照正文引用顺序排序
\bibliography{biblio}% Produces the bibliography via BibTeX.

@incollection{landau2013statistical,
title = {CHAPTER II - THERMODYNAMIC QUANTITIES},
editor = {L.D. LANDAU and E.M. LIFSHITZ},
booktitle = {Statistical Physics (Third Edition)},
publisher = {Butterworth-Heinemann},
edition = {Third Edition},
address = {Oxford},
pages = {34-78},
year = {1980},
isbn = {978-0-08-057046-4},
doi = {https://doi.org/10.1016/B978-0-08-057046-4.50009-9},
url = {https://www.sciencedirect.com/science/article/pii/B9780080570464500099},
author = {L.D. LANDAU and E.M. LIFSHITZ}
}

@article{PhysRevLett.126.161101,
  title = {Exoplanets as Sub-GeV Dark Matter Detectors},
  author = {Leane, Rebecca K. and Smirnov, Juri},
  journal = {Phys. Rev. Lett.},
  volume = {126},
  issue = {16},
  pages = {161101},
  numpages = {11},
  year = {2021},
  month = {Apr},
  publisher = {American Physical Society},
  doi = {10.1103/PhysRevLett.126.161101},
  url = {https://link.aps.org/doi/10.1103/PhysRevLett.126.161101}
}

@article{10.1093/mnras/stz1698,
    author = {Lin, Hai-Nan and Li, Xin},
    title = {The dark matter profiles in the Milky Way},
    journal = {Monthly Notices of the Royal Astronomical Society},
    volume = {487},
    number = {4},
    pages = {5679-5684},
    year = {2019},
    month = {06},
    issn = {0035-8711},
    doi = {10.1093/mnras/stz1698},
    url = {https://doi.org/10.1093/mnras/stz1698},
    eprint = {https://academic.oup.com/mnras/article-pdf/487/4/5679/28897927/stz1698.pdf},
}

@article{Planck:2018vyg,
    author = "Aghanim, N. and others",
    collaboration = "Planck",
    title = "{Planck 2018 results. VI. Cosmological parameters}",
    eprint = "1807.06209",
    archivePrefix = "arXiv",
    primaryClass = "astro-ph.CO",
    doi = "10.1051/0004-6361/201833910",
    journal = "Astron. Astrophys.",
    volume = "641",
    pages = "A6",
    year = "2020",
    note = "[Erratum: Astron.Astrophys. 652, C4 (2021)]"
}

@article{Bertone:2004pz,
    author = "Bertone, Gianfranco and Hooper, Dan and Silk, Joseph",
    title = "{Particle dark matter: Evidence, candidates and constraints}",
    eprint = "hep-ph/0404175",
    archivePrefix = "arXiv",
    reportNumber = "FERMILAB-PUB-04-047-A",
    doi = "10.1016/j.physrep.2004.08.031",
    journal = "Phys. Rept.",
    volume = "405",
    pages = "279--390",
    year = "2005"
}

@article{Young:2016ala,
    author = "Young, Bing-Lin",
    title = "{A survey of dark matter and related topics in cosmology}",
    doi = "10.1007/s11467-016-0583-4",
    journal = "Front. Phys. (Beijing)",
    volume = "12",
    number = "2",
    pages = "121201",
    year = "2017",
    note = "[Erratum: Front.Phys.(Beijing) 12, 121202 (2017)]"
}

@article{Billard:2021uyg,
    author = "Billard, Julien and others",
    title = "{Direct detection of dark matter\textemdash{}APPEC committee report*}",
    eprint = "2104.07634",
    archivePrefix = "arXiv",
    primaryClass = "hep-ex",
    doi = "10.1088/1361-6633/ac5754",
    journal = "Rept. Prog. Phys.",
    volume = "85",
    number = "5",
    pages = "056201",
    year = "2022"
}

@article{Hui:2021tkt,
    author = "Hui, Lam",
    title = "{Wave Dark Matter}",
    eprint = "2101.11735",
    archivePrefix = "arXiv",
    primaryClass = "astro-ph.CO",
    doi = "10.1146/annurev-astro-120920-010024",
    journal = "Ann. Rev. Astron. Astrophys.",
    volume = "59",
    pages = "247--289",
    year = "2021"
}

@article{Wilczek:1977pj,
    author = "Wilczek, Frank",
    title = "{Problem of Strong  $P$  and  $T$  Invariance in the Presence of Instantons}",
    reportNumber = "Print-77-0939 (COLUMBIA)",
    doi = "10.1103/PhysRevLett.40.279",
    journal = "Phys. Rev. Lett.",
    volume = "40",
    pages = "279--282",
    year = "1978"
}

@article{Svrcek:2006yi,
    author = "Svrcek, Peter and Witten, Edward",
    title = "{Axions In String Theory}",
    eprint = "hep-th/0605206",
    archivePrefix = "arXiv",
    reportNumber = "SLAC-PUB-11894",
    doi = "10.1088/1126-6708/2006/06/051",
    journal = "JHEP",
    volume = "06",
    pages = "051",
    year = "2006"
}

@article{PRESKILL1983127,
title = {Cosmology of the invisible axion},
journal = {Physics Letters B},
volume = {120},
number = {1},
pages = {127-132},
year = {1983},
issn = {0370-2693},
doi = {https://doi.org/10.1016/0370-2693(83)90637-8},
url = {https://www.sciencedirect.com/science/article/pii/0370269383906378},
author = {John Preskill and Mark B. Wise and Frank Wilczek}
}

@article{Lacki_2010,
doi = {10.1088/2041-8205/720/1/L67},
url = {https://dx.doi.org/10.1088/2041-8205/720/1/L67},
year = {2010},
month = {aug},
publisher = {The American Astronomical Society},
volume = {720},
number = {1},
pages = {L67},
author = {Lacki, Brian C. and Beacom, John F.},
title = {PRIMORDIAL BLACK HOLES AS DARK MATTER: ALMOST ALL OR ALMOST NOTHING},
journal = {The Astrophysical Journal Letters}
}

@article{Hu:2000ke,
    author = "Hu, Wayne and Barkana, Rennan and Gruzinov, Andrei",
    title = "{Cold and fuzzy dark matter}",
    eprint = "astro-ph/0003365",
    archivePrefix = "arXiv",
    doi = "10.1103/PhysRevLett.85.1158",
    journal = "Phys. Rev. Lett.",
    volume = "85",
    pages = "1158--1161",
    year = "2000"
}

@article{Burkert:2020laq,
    author = "Burkert, Andreas",
    title = "{Fuzzy Dark Matter and Dark Matter Halo Cores}",
    eprint = "2006.11111",
    archivePrefix = "arXiv",
    primaryClass = "astro-ph.GA",
    doi = "10.3847/1538-4357/abb242",
    journal = "Astrophys. J.",
    volume = "904",
    number = "2",
    pages = "161",
    year = "2020"
}

@article{Carr:1974nx,
    author = "Carr, Bernard J. and Hawking, S. W.",
    title = "{Black holes in the early Universe}",
    doi = "10.1093/mnras/168.2.399",
    journal = "Mon. Not. Roy. Astron. Soc.",
    volume = "168",
    pages = "399--415",
    year = "1974"
}

@article{Carr:2016drx,
    author = "Carr, Bernard and Kuhnel, Florian and Sandstad, Marit",
    title = "{Primordial Black Holes as Dark Matter}",
    eprint = "1607.06077",
    archivePrefix = "arXiv",
    primaryClass = "astro-ph.CO",
    reportNumber = "NORDITA-2016-83",
    doi = "10.1103/PhysRevD.94.083504",
    journal = "Phys. Rev. D",
    volume = "94",
    number = "8",
    pages = "083504",
    year = "2016"
}

@article{PhysRevD.76.043523,
  title = {Towards closing the window on strongly interacting dark matter: Far-reaching constraints from Earth's heat flow},
  author = {Mack, Gregory D. and Beacom, John F. and Bertone, Gianfranco},
  journal = {Phys. Rev. D},
  volume = {76},
  issue = {4},
  pages = {043523},
  numpages = {12},
  year = {2007},
  month = {Aug},
  publisher = {American Physical Society},
  doi = {10.1103/PhysRevD.76.043523},
  url = {https://link.aps.org/doi/10.1103/PhysRevD.76.043523}
}

@article{PhysRevD.94.035024,
  title = {Constraints on leptophilic light dark matter from internal heat flux of Earth},
  author = {Chauhan, Bhavesh and Mohanty, Subhendra},
  journal = {Phys. Rev. D},
  volume = {94},
  issue = {3},
  pages = {035024},
  numpages = {6},
  year = {2016},
  month = {Aug},
  publisher = {American Physical Society},
  doi = {10.1103/PhysRevD.94.035024},
  url = {https://link.aps.org/doi/10.1103/PhysRevD.94.035024}
}

@article{PhysRevD.101.043001,
  title = {Terrestrial and martian heat flow limits on dark matter},
  author = {Bramante, Joseph and Buchanan, Andrew and Goodman, Alan and Lodhi, Eesha},
  journal = {Phys. Rev. D},
  volume = {101},
  issue = {4},
  pages = {043001},
  numpages = {14},
  year = {2020},
  month = {Feb},
  publisher = {American Physical Society},
  doi = {10.1103/PhysRevD.101.043001},
  url = {https://link.aps.org/doi/10.1103/PhysRevD.101.043001}
}

@misc{feng2018detectionclosestjovianexoplanet,
      title={Detection of the closest Jovian exoplanet in the Epsilon Indi triple system}, 
      author={Fabo Feng and Mikko Tuomi and Hugh R. A. Jones},
      year={2018},
      eprint={1803.08163},
      archivePrefix={arXiv},
      primaryClass={astro-ph.EP},
      url={https://arxiv.org/abs/1803.08163}, 
}

@article{Bailey_2009,
doi = {10.1088/0004-637X/690/1/743},
url = {https://dx.doi.org/10.1088/0004-637X/690/1/743},
year = {2008},
month = {dec},
publisher = {The American Astronomical Society},
volume = {690},
number = {1},
pages = {743},
author = {Bailey, Jeremy and Butler, R. Paul and Tinney, C. G. and Jones, Hugh R. A. and O'Toole, Simon and Carter, Brad D. and Marcy, Geoffrey W.},
title = {A JUPITER-LIKE PLANET ORBITING THE NEARBY M DWARF GJ 832*},
journal = {The Astrophysical Journal}
}

@article{Butler_2006,
doi = {10.1086/510500},
url = {https://dx.doi.org/10.1086/510500},
year = {2006},
month = {nov},
publisher = {The University of Chicago Press},
volume = {118},
number = {850},
pages = {1685},
author = {Butler, R. Paul and Johnson, John Asher and Marcy, Geoffrey W. and Wright, Jason T. and Vogt, Steven S. and Fischer, Debra A.},
title = {A Long‐Period Jupiter‐Mass Planet Orbiting the Nearby M Dwarf GJ 8491},
journal = {Publications of the Astronomical Society of the Pacific}
}

@article{Marcy_2002,
doi = {10.1086/344298},
url = {https://dx.doi.org/10.1086/344298},
year = {2002},
month = {dec},
publisher = {},
volume = {581},
number = {2},
pages = {1375},
author = {Marcy, Geoffrey W. and Butler, R. Paul and Fischer, Debra A. and Laughlin, Greg and Vogt, Steven S. and Henry, Gregory W. and Pourbaix, Dimitri},
title = {A Planet at 5 AU around 55 Cancri*},
journal = {The Astrophysical Journal}
}

@article{10.1111/j.1365-2966.2009.16233.x,
    author = {Gregory, Philip C. and Fischer, Debra A.},
    title = {A Bayesian periodogram finds evidence for three planets in 47 Ursae Majoris},
    journal = {Monthly Notices of the Royal Astronomical Society},
    volume = {403},
    number = {2},
    pages = {731-747},
    year = {2010},
    month = {03},
    issn = {0035-8711},
    doi = {10.1111/j.1365-2966.2009.16233.x},
    url = {https://doi.org/10.1111/j.1365-2966.2009.16233.x},
    eprint = {https://academic.oup.com/mnras/article-pdf/403/2/731/4002836/mnras0403-0731.pdf},
}

@article{Johnson_2007,
doi = {10.1086/521720},
url = {https://dx.doi.org/10.1086/521720},
year = {2007},
month = {nov},
publisher = {},
volume = {670},
number = {1},
pages = {833},
author = {Johnson, John Asher and Butler, R. Paul and Marcy, Geoffrey W. and Fischer, Debra A. and Vogt, Steven S. and Wright, Jason T. and Peek, Kathryn M. G.},
title = {A New Planet around an M Dwarf: Revealing a Correlation between Exoplanets and Stellar Mass*},
journal = {The Astrophysical Journal}
}

@article{Fischer_2009,
doi = {10.1088/0004-637X/703/2/1545},
url = {https://dx.doi.org/10.1088/0004-637X/703/2/1545},
year = {2009},
month = {sep},
publisher = {The American Astronomical Society},
volume = {703},
number = {2},
pages = {1545},
author = {Fischer, Debra and Driscoll, Peter and Isaacson, Howard and Giguere, Matt and Marcy, Geoffrey W. and Valenti, Jeff and Wright, Jason T. and Henry, Gregory W. and Johnson, John Asher and Howard, Andrew and Peek, Katherine and McCarthy, Chris},
title = {FIVE PLANETS AND AN INDEPENDENT CONFIRMATION OF HD 196885Ab FROM LICK OBSERVATORY*},
journal = {The Astrophysical Journal}
}

@article{ Rey2017,
	author = {{Rey, J.} and {Hébrard, G.} and {Bouchy, F.} and {Bourrier, V.} and {Boisse, I.} and {Santos, N. C.} and {Arnold, L.} and {Astudillo-Defru, N.} and {Bonfils, X.} and {Borgniet, S.} and {Courcol, B.} and {Deleuil, M.} and {Delfosse, X.} and {Demangeon, O.} and {Díaz, R. F.} and {Ehrenreich, D.} and {Forveille, T.} and {Marmier, M.} and {Moutou, C.} and {Pepe, F.} and {Santerne, A.} and {Sahlmann, J.} and {Ségransan, D.} and {Udry, S.} and {Wilson, P. A.}},
	title = {The SOPHIE search for northern extrasolar planets - XII. Three giant planets suitable for astrometric mass determination  with Gaia⋆⋆⋆},
	DOI= "10.1051/0004-6361/201630089",
	url= "https://doi.org/10.1051/0004-6361/201630089",
	journal = {A\&A},
	year = 2017,
	volume = 601,
	pages = "A9",
}

@article{Yee_2018,
doi = {10.3847/1538-3881/aabfec},
url = {https://dx.doi.org/10.3847/1538-3881/aabfec},
year = {2018},
month = {may},
publisher = {The American Astronomical Society},
volume = {155},
number = {6},
pages = {255},
author = {Yee, Samuel W. and Petigura, Erik A. and Fulton, Benjamin J. and Knutson, Heather A. and Batygin, Konstantin and Bakos, G{\'a}sp{\'a}r {\'A}. and Hartman, Joel D. and Hirsch, Lea A. and Howard, Andrew W. and Isaacson, Howard and Kosiarek, Molly R. and Sinukoff, Evan and Weiss, Lauren M.},
title = {HAT-P-11: Discovery of a Second Planet and a Clue to Understanding Exoplanet Obliquities},
journal = {The Astronomical Journal}
}

@article{Vogt_2002,
doi = {10.1086/338768},
url = {https://dx.doi.org/10.1086/338768},
year = {2002},
month = {mar},
publisher = {},
volume = {568},
number = {1},
pages = {352},
author = {Vogt, Steven S. and Butler, R. Paul and Marcy, Geoffrey W. and Fischer, Debra A. and Pourbaix, Dimitri and Apps, Kevin and Laughlin, Gregory},
title = {Ten Low-Mass Companions from the Keck Precision Velocity Survey*},
journal = {The Astrophysical Journal}
}

@article{Niedzielski_2009,
doi = {10.1088/0004-637X/707/1/768},
url = {https://dx.doi.org/10.1088/0004-637X/707/1/768},
year = {2009},
month = {nov},
publisher = {The American Astronomical Society},
volume = {707},
number = {1},
pages = {768},
author = {Niedzielski, A. and Nowak, G. and Adamów, M. and Wolszczan, A.},
title = {SUBSTELLAR-MASS COMPANIONS TO THE K-DWARF BD+14 4559 AND THE K-GIANTS HD 240210 AND BD+20 2457},
journal = {The Astrophysical Journal}
}

@article{Marmier2013,
	author = {{Marmier, M.} , {Ségransan, D.} and et al.}, 
	title = {The CORALIE 
survey for southern extrasolar 
            planets - XVII. New and updated long period and massive planets⋆⋆⋆},
	DOI= "10.1051/0004-6361/201219639",
	url= "https://doi.org/10.1051/0004-6361/201219639",
	journal = {A\&A},
	year = 2013,
	volume = 551,
	pages = "A90",
	month = "",
}

@article{ Naef2010,
	author = {{Naef, D.} , {Mayor, M.} and et al.},
	title = {The HARPS search for southern extrasolar planets ⋆⋆⋆ - XXIII. 8 planetary companions to low-activity solar-type stars},
	DOI= "10.1051/0004-6361/200913616",
	url= "https://doi.org/10.1051/0004-6361/200913616",
	journal = {A\&A},
	year = 2010,
	volume = 523,
	pages = "A15",
	month = "",
}

@article{PhysRevD.40.3221,
  title = {Weakly interacting massive particles and neutron stars},
  author = {Goldman, Itzhak and Nussinov, Shmuel},
  journal = {Phys. Rev. D},
  volume = {40},
  issue = {10},
  pages = {3221--3230},
  numpages = {0},
  year = {1989},
  month = {Nov},
  publisher = {American Physical Society},
  doi = {10.1103/PhysRevD.40.3221},
  url = {https://link.aps.org/doi/10.1103/PhysRevD.40.3221}
}

@article{Gould:1989gw,
    author = "Gould, Andrew and Draine, Bruce T. and Romani, Roger W. and Nussinov, Shmuel",
    title = "{Neutron Stars: Graveyard of Charged Dark Matter}",
    reportNumber = "IASSNS-AST 89/55",
    doi = "10.1016/0370-2693(90)91745-W",
    journal = "Phys. Lett. B",
    volume = "238",
    pages = "337--343",
    year = "1990"
}

@article{PhysRevD.77.023006,
  title = {WIMP annihilation and cooling of neutron stars},
  author = {Kouvaris, Chris},
  journal = {Phys. Rev. D},
  volume = {77},
  issue = {2},
  pages = {023006},
  numpages = {9},
  year = {2008},
  month = {Jan},
  publisher = {American Physical Society},
  doi = {10.1103/PhysRevD.77.023006},
  url = {https://link.aps.org/doi/10.1103/PhysRevD.77.023006}
}

@article{PhysRevD.77.043515,
  title = {Compact stars as dark matter probes},
  author = {Bertone, Gianfranco and Fairbairn, Malcolm},
  journal = {Phys. Rev. D},
  volume = {77},
  issue = {4},
  pages = {043515},
  numpages = {9},
  year = {2008},
  month = {Feb},
  publisher = {American Physical Society},
  doi = {10.1103/PhysRevD.77.043515},
  url = {https://link.aps.org/doi/10.1103/PhysRevD.77.043515}
}

@article{PhysRevD.81.123521,
  title = {Neutron stars as dark matter probes},
  author = {de Lavallaz, Arnaud and Fairbairn, Malcolm},
  journal = {Phys. Rev. D},
  volume = {81},
  issue = {12},
  pages = {123521},
  numpages = {10},
  year = {2010},
  month = {Jun},
  publisher = {American Physical Society},
  doi = {10.1103/PhysRevD.81.123521},
  url = {https://link.aps.org/doi/10.1103/PhysRevD.81.123521}
}

@article{PhysRevD.82.063531,
  title = {Can neutron stars constrain dark matter?},
  author = {Kouvaris, Chris and Tinyakov, Peter},
  journal = {Phys. Rev. D},
  volume = {82},
  issue = {6},
  pages = {063531},
  numpages = {9},
  year = {2010},
  month = {Sep},
  publisher = {American Physical Society},
  doi = {10.1103/PhysRevD.82.063531},
  url = {https://link.aps.org/doi/10.1103/PhysRevD.82.063531}
}

@article{PhysRevD.85.023519,
  title = {Constraints on scalar asymmetric dark matter from black hole formation in neutron stars},
  author = {McDermott, Samuel D. and Yu, Hai-Bo and Zurek, Kathryn M.},
  journal = {Phys. Rev. D},
  volume = {85},
  issue = {2},
  pages = {023519},
  numpages = {12},
  year = {2012},
  month = {Jan},
  publisher = {American Physical Society},
  doi = {10.1103/PhysRevD.85.023519},
  url = {https://link.aps.org/doi/10.1103/PhysRevD.85.023519}
}

@article{PhysRevLett.107.091301,
  title = {Excluding Light Asymmetric Bosonic Dark Matter},
  author = {Kouvaris, Chris and Tinyakov, Peter},
  journal = {Phys. Rev. Lett.},
  volume = {107},
  issue = {9},
  pages = {091301},
  numpages = {4},
  year = {2011},
  month = {Aug},
  publisher = {American Physical Society},
  doi = {10.1103/PhysRevLett.107.091301},
  url = {https://link.aps.org/doi/10.1103/PhysRevLett.107.091301}
}

@article{Guver_2014,
doi = {10.1088/1475-7516/2014/05/013},
url = {https://dx.doi.org/10.1088/1475-7516/2014/05/013},
year = {2014},
month = {may},
publisher = {},
volume = {2014},
number = {05},
pages = {013},
author = {Güver, Tolga and Erkoca, Arif Emre and Reno, Mary Hall and Sarcevic, Ina},
title = {On the  capture of dark matter by  neutron stars},
journal = {Journal of Cosmology and Astroparticle Physics}
}

@article{PhysRevD.87.055012,
  title = {Constraints on bosonic dark matter from observation of old neutron stars},
  author = {Bramante, Joseph and Fukushima, Keita and Kumar, Jason},
  journal = {Phys. Rev. D},
  volume = {87},
  issue = {5},
  pages = {055012},
  numpages = {9},
  year = {2013},
  month = {Mar},
  publisher = {American Physical Society},
  doi = {10.1103/PhysRevD.87.055012},
  url = {https://link.aps.org/doi/10.1103/PhysRevD.87.055012}
}

@article{PhysRevD.87.123507,
  title = {Realistic neutron star constraints on bosonic asymmetric dark matter},
  author = {Bell, Nicole F. and Melatos, Andrew and Petraki, Kalliopi},
  journal = {Phys. Rev. D},
  volume = {87},
  issue = {12},
  pages = {123507},
  numpages = {15},
  year = {2013},
  month = {Jun},
  publisher = {American Physical Society},
  doi = {10.1103/PhysRevD.87.123507},
  url = {https://link.aps.org/doi/10.1103/PhysRevD.87.123507}
}

@article{PhysRevD.88.123505,
  title = {Dark matter thermalization in neutron stars},
  author = {Bertoni, Bridget and Nelson, Ann E. and Reddy, Sanjay},
  journal = {Phys. Rev. D},
  volume = {88},
  issue = {12},
  pages = {123505},
  numpages = {11},
  year = {2013},
  month = {Dec},
  publisher = {American Physical Society},
  doi = {10.1103/PhysRevD.88.123505},
  url = {https://link.aps.org/doi/10.1103/PhysRevD.88.123505}
}

@article{PhysRevD.83.083512,
  title = {Constraining asymmetric dark matter through observations of compact stars},
  author = {Kouvaris, Chris and Tinyakov, Peter},
  journal = {Phys. Rev. D},
  volume = {83},
  issue = {8},
  pages = {083512},
  numpages = {9},
  year = {2011},
  month = {Apr},
  publisher = {American Physical Society},
  doi = {10.1103/PhysRevD.83.083512},
  url = {https://link.aps.org/doi/10.1103/PhysRevD.83.083512}
}

@article{PhysRevD.81.083520,
  title = {Capture of inelastic dark matter in white dwarves},
  author = {McCullough, Matthew and Fairbairn, Malcolm},
  journal = {Phys. Rev. D},
  volume = {81},
  issue = {8},
  pages = {083520},
  numpages = {8},
  year = {2010},
  month = {Apr},
  publisher = {American Physical Society},
  doi = {10.1103/PhysRevD.81.083520},
  url = {https://link.aps.org/doi/10.1103/PhysRevD.81.083520}
}

@article{PhysRevLett.115.141301,
  title = {Dark Matter Ignition of Type Ia Supernovae},
  author = {Bramante, Joseph},
  journal = {Phys. Rev. Lett.},
  volume = {115},
  issue = {14},
  pages = {141301},
  numpages = {6},
  year = {2015},
  month = {Sep},
  publisher = {American Physical Society},
  doi = {10.1103/PhysRevLett.115.141301},
  url = {https://link.aps.org/doi/10.1103/PhysRevLett.115.141301}
}

@article{PhysRevD.92.063007,
  title = {Dark matter triggers of supernovae},
  author = {Graham, Peter W. and Rajendran, Surjeet and Varela, Jaime},
  journal = {Phys. Rev. D},
  volume = {92},
  issue = {6},
  pages = {063007},
  numpages = {7},
  year = {2015},
  month = {Sep},
  publisher = {American Physical Society},
  doi = {10.1103/PhysRevD.92.063007},
  url = {https://link.aps.org/doi/10.1103/PhysRevD.92.063007}
}

@article{PhysRevD.94.063001,
  title = {Light dark matter scattering in outer neutron star crusts},
  author = {Cerme\~no, Marina and P\'erez-Garc\'{\i}a, M.\'Angeles and Silk, Joseph},
  journal = {Phys. Rev. D},
  volume = {94},
  issue = {6},
  pages = {063001},
  numpages = {7},
  year = {2016},
  month = {Sep},
  publisher = {American Physical Society},
  doi = {10.1103/PhysRevD.94.063001},
  url = {https://link.aps.org/doi/10.1103/PhysRevD.94.063001}
}

@article{PhysRevD.98.115027,
  title = {White dwarfs as dark matter detectors},
  author = {Graham, Peter W. and Janish, Ryan and Narayan, Vijay and Rajendran, Surjeet and Riggins, Paul},
  journal = {Phys. Rev. D},
  volume = {98},
  issue = {11},
  pages = {115027},
  numpages = {20},
  year = {2018},
  month = {Dec},
  publisher = {American Physical Society},
  doi = {10.1103/PhysRevD.98.115027},
  url = {https://link.aps.org/doi/10.1103/PhysRevD.98.115027}
}

@article{PhysRevD.100.043020,
  title = {Supernovae sparked by dark matter in white dwarfs},
  author = {Acevedo, Javier F. and Bramante, Joseph},
  journal = {Phys. Rev. D},
  volume = {100},
  issue = {4},
  pages = {043020},
  numpages = {20},
  year = {2019},
  month = {Aug},
  publisher = {American Physical Society},
  doi = {10.1103/PhysRevD.100.043020},
  url = {https://link.aps.org/doi/10.1103/PhysRevD.100.043020}
}

@article{Krall_2018,
doi = {10.1088/1674-1137/42/4/043105},
url = {https://dx.doi.org/10.1088/1674-1137/42/4/043105},
year = {2018},
month = {apr},
publisher = {Chinese Physical Society and the Institute of High Energy Physics of the Chinese Academy of Sciences and the Institute of Modern Physics of the Chinese Academy of Sciences and IOP Publishing Ltd},
volume = {42},
number = {4},
pages = {043105},
author = {Krall, Rebecca and Reece, Matthew},
title = {Last electroweak WIMP standing: pseudo-dirac higgsino status and compact stars as future probes*},
journal = {Chinese Physics C}
}

@article{Dasgupta_2019,
doi = {10.1088/1475-7516/2019/08/018},
url = {https://dx.doi.org/10.1088/1475-7516/2019/08/018},
year = {2019},
month = {aug},
publisher = {},
volume = {2019},
number = {08},
pages = {018},
author = {Dasgupta, Basudeb and Gupta, Aritra and Ray, Anupam},
title = {Dark matter capture in celestial objects: improved treatment of multiple scattering and updated constraints from white dwarfs},
journal = {Journal of Cosmology and Astroparticle Physics}
}

@article{PhysRevD.70.103517,
  title = {Uranus's anomalously low excess heat constrains strongly interacting dark matter},
  author = {Mitra, Saibal},
  journal = {Phys. Rev. D},
  volume = {70},
  issue = {10},
  pages = {103517},
  numpages = {6},
  year = {2004},
  month = {Nov},
  publisher = {American Physical Society},
  doi = {10.1103/PhysRevD.70.103517},
  url = {https://link.aps.org/doi/10.1103/PhysRevD.70.103517}
}

@article{10.1143/ptp/87.3.685,
    author = {Kawasaki, Masahiro and Murayama, Hitoshi and Yanagida, Tsutomu},
    title = {Can the Strongly Interacting Dark Matter Be a Heating Source of Jupiter?},
    journal = {Progress of Theoretical Physics},
    volume = {87},
    number = {3},
    pages = {685-692},
    year = {1992},
    month = {03},
    issn = {0033-068X},
    doi = {10.1143/ptp/87.3.685},
    url = {https://doi.org/10.1143/ptp/87.3.685},
    eprint = {https://academic.oup.com/ptp/article-pdf/87/3/685/5240586/87-3-685.pdf},
}

@article{Leane:2020wob,
    author = "Leane, Rebecca K. and Smirnov, Juri",
    title = "{Exoplanets as Sub-GeV Dark Matter Detectors}",
    eprint = "2010.00015",
    archivePrefix = "arXiv",
    primaryClass = "hep-ph",
    reportNumber = "MIT-CTP/5230, SLAC-PUB-17556",
    doi = "10.1103/PhysRevLett.126.161101",
    journal = "Phys. Rev. Lett.",
    volume = "126",
    number = "16",
    pages = "161101",
    year = "2021"
}

@article{Blanco:2023qgi,
    author = "Blanco, Carlos and Leane, Rebecca K.",
    title = "{Search for Dark Matter Ionization on the Night Side of Jupiter with Cassini}",
    eprint = "2312.06758",
    archivePrefix = "arXiv",
    primaryClass = "hep-ph",
    reportNumber = "SLAC-PUB-17753",
    doi = "10.1103/PhysRevLett.132.261002",
    journal = "Phys. Rev. Lett.",
    volume = "132",
    number = "26",
    pages = "261002",
    year = "2024"
}

@article{Benito:2024yki,
    author = "Benito, Mar\'\i{}a and Karchev, Konstantin and Leane, Rebecca K. and P\~oder, Sven and Smirnov, Juri and Trotta, Roberto",
    title = "{Dark Matter halo parameters from overheated exoplanets via Bayesian hierarchical inference}",
    eprint = "2405.09578",
    archivePrefix = "arXiv",
    primaryClass = "astro-ph.IM",
    doi = "10.1088/1475-7516/2024/07/038",
    journal = "JCAP",
    volume = "07",
    pages = "038",
    year = "2024"
}

@article{Stephens:2021sfq,
    author = "Stephens, Marric",
    title = "{Detecting Dark Matter in Exoplanets}",
    doi = "10.1103/Physics.14.s46",
    journal = "APS Physics",
    volume = "14",
    pages = "s46",
    year = "2021"
}

@ARTICLE{Phoroutan-Mehr:2025hjz,
       author = {{Phoroutan-Mehr}, Mehrdad and {Fetherolf}, Tara},
        title = "{Probing Superheavy Dark Matter with Exoplanets}",
      journal = {arXiv e-prints},
         year = 2025,
        month = feb,
          eid = {arXiv:2503.00125},
        pages = {arXiv:2503.00125},
          doi = {10.48550/arXiv.2503.00125},
archivePrefix = {arXiv},
       eprint = {2503.00125},
 primaryClass = {hep-ph},
       adsurl = {https://ui.adsabs.harvard.edu/abs/2025arXiv250300125P}
}

@ARTICLE{2015AA...577A..42B,
       author = {{Baraffe}, Isabelle and {Homeier}, Derek and {Allard}, France and {Chabrier}, Gilles},
        title = "{New evolutionary models for pre-main sequence and main sequence low-mass stars down to the hydrogen-burning limit}",
      journal = {Astronomy \& Astrophysics},
         year = 2015,
        month = may,
       volume = {577},
          eid = {A42},
        pages = {A42},
          doi = {10.1051/0004-6361/201425481},
archivePrefix = {arXiv},
       eprint = {1503.04107},
 primaryClass = {astro-ph.SR},
       adsurl = {https://ui.adsabs.harvard.edu/abs/2015A\&A...577A..42B}
}

@article{Baines_2012,
doi = {10.1088/0004-637X/748/1/72},
url = {https://dx.doi.org/10.1088/0004-637X/748/1/72},
year = {2012},
month = {mar},
publisher = {The American Astronomical Society},
volume = {748},
number = {1},
pages = {72},
author = {Baines, Ellyn K. and Thomas Armstrong, J.},
title = {ERRATUM: "CONFIRMING FUNDAMENTAL PROPERTIES OF THE EXOPLANET HOST STAR ϵ ERIDANI USING THE NAVY OPTICAL INTERFEROMETER" (2012, ApJ, 744, 138B)},
journal = {The Astrophysical Journal}
}

@article{Adler:2008ky,
    author = "Adler, Stephen L.",
    title = "{Planet-bound dark matter and the internal heat of Uranus, Neptune, and hot-Jupiter exoplanets}",
    eprint = "0808.2823",
    archivePrefix = "arXiv",
    primaryClass = "astro-ph",
    doi = "10.1016/j.physletb.2008.12.023",
    journal = "Phys. Lett. B",
    volume = "671",
    pages = "203--206",
    year = "2009"
}

@article{Bramante:2022pmn,
    author = "Bramante, Joseph and Kumar, Jason and Mohlabeng, Gopolang and Raj, Nirmal and Song, Ningqiang",
    title = "{Light dark matter accumulating in planets: Nuclear scattering}",
    eprint = "2210.01812",
    archivePrefix = "arXiv",
    primaryClass = "hep-ph",
    reportNumber = "UCI-HEP-TR-2022-18",
    doi = "10.1103/PhysRevD.108.063022",
    journal = "Phys. Rev. D",
    volume = "108",
    number = "6",
    pages = "063022",
    year = "2023"
}

@article{Chauhan:2016joa,
    author = "Chauhan, Bhavesh and Mohanty, Subhendra",
    title = "{Constraints on leptophilic light dark matter from internal heat flux of Earth}",
    eprint = "1603.06350",
    archivePrefix = "arXiv",
    primaryClass = "hep-ph",
    doi = "10.1103/PhysRevD.94.035024",
    journal = "Phys. Rev. D",
    volume = "94",
    number = "3",
    pages = "035024",
    year = "2016"
}

@article{Gardner:2006ky,
    author = "Gardner, Jonathan P. and others",
    title = "{The James Webb Space Telescope}",
    eprint = "astro-ph/0606175",
    archivePrefix = "arXiv",
    doi = "10.1007/s11214-006-8315-7",
    journal = "Space Sci. Rev.",
    volume = "123",
    pages = "485",
    year = "2006"
}

@article{Baryakhtar:2017dbj,
    author = "Baryakhtar, Masha and Bramante, Joseph and Li, Shirley Weishi and Linden, Tim and Raj, Nirmal",
    title = "{Dark Kinetic Heating of Neutron Stars and An Infrared Window On WIMPs, SIMPs, and Pure Higgsinos}",
    eprint = "1704.01577",
    archivePrefix = "arXiv",
    primaryClass = "hep-ph",
    doi = "10.1103/PhysRevLett.119.131801",
    journal = "Phys. Rev. Lett.",
    volume = "119",
    number = "13",
    pages = "131801",
    year = "2017"
}

@article{Deming:2009is,
    author = "Deming, D. and others",
    title = "{Discovery and Characterization of Transiting SuperEarths Using an All-Sky Transit Survey and Follow-up by the James Webb Space Telescope}",
    eprint = "0903.4880",
    archivePrefix = "arXiv",
    primaryClass = "astro-ph.EP",
    doi = "10.1086/605913",
    journal = "Publ. Astron. Soc. Pac.",
    volume = "121",
    pages = "952",
    year = "2009"
}

@article{Smith:2019wgb,
    author = "Smith, Krista Lynne",
    title = "{The vast potential of exoplanet satellites for high-energy astrophysics}",
    eprint = "1904.08952",
    archivePrefix = "arXiv",
    primaryClass = "astro-ph.IM",
    doi = "10.1002/asna.201913615",
    journal = "Astron. Nachr.",
    volume = "340",
    number = "4",
    pages = "308--313",
    year = "2019"
}

@ARTICLE{2024arXiv241012905W,
       author = {{Winn}, Joshua N.},
        title = "{The Transiting Exoplanet Survey Satellite}",
      journal = {arXiv e-prints},
         year = 2024,
        month = oct,
          eid = {arXiv:2410.12905},
        pages = {arXiv:2410.12905},
          doi = {10.48550/arXiv.2410.12905},
archivePrefix = {arXiv},
       eprint = {2410.12905},
 primaryClass = {astro-ph.EP},
       adsurl = {https://ui.adsabs.harvard.edu/abs/2024arXiv241012905W}
}

@article{deSalas_2021,
doi = {10.1088/1361-6633/ac24e7},
url = {https://dx.doi.org/10.1088/1361-6633/ac24e7},
year = {2021},
month = {oct},
publisher = {IOP Publishing},
volume = {84},
number = {10},
pages = {104901},
author = {de Salas, Pablo F and Widmark, A},
title = {Dark matter local density determination: recent observations and future prospects},
journal = {Reports on Progress in Physics}
}

@article{Hooper:2011dw,
    author = "Hooper, Dan and Steffen, Jason H.",
    title = "{Dark Matter And The Habitability of Planets}",
    eprint = "1103.5086",
    archivePrefix = "arXiv",
    primaryClass = "astro-ph.EP",
    reportNumber = "FERMILAB-PUB-11-149-A",
    doi = "10.1088/1475-7516/2012/07/046",
    journal = "JCAP",
    volume = "07",
    pages = "046",
    year = "2012"
}

@article{Safronova:2017xyt,
    author = "Safronova, M. S. and Budker, D. and DeMille, D. and Kimball, Derek F. Jackson and Derevianko, A. and Clark, C. W.",
    title = "{Search for New Physics with Atoms and Molecules}",
    eprint = "1710.01833",
    archivePrefix = "arXiv",
    primaryClass = "physics.atom-ph",
    doi = "10.1103/RevModPhys.90.025008",
    journal = "Rev. Mod. Phys.",
    volume = "90",
    number = "2",
    pages = "025008",
    year = "2018"
}

@article{Alekhin:2015byh,
    author = "Alekhin, Sergey and others",
    title = "{A facility to Search for Hidden Particles at the CERN SPS: the SHiP physics case}",
    eprint = "1504.04855",
    archivePrefix = "arXiv",
    primaryClass = "hep-ph",
    reportNumber = "CERN-SPSC-2015-017, SPSC-P-350-ADD-1",
    doi = "10.1088/0034-4885/79/12/124201",
    journal = "Rept. Prog. Phys.",
    volume = "79",
    number = "12",
    pages = "124201",
    year = "2016"
}

@article{ADMX:2009iij,
    author = "Asztalos, S. J. and others",
    collaboration = "ADMX",
    title = "{A SQUID-based microwave cavity search for dark-matter axions}",
    eprint = "0910.5914",
    archivePrefix = "arXiv",
    primaryClass = "astro-ph.CO",
    doi = "10.1103/PhysRevLett.104.041301",
    journal = "Phys. Rev. Lett.",
    volume = "104",
    pages = "041301",
    year = "2010"
}

@article{Goodman:2010ku,
    author = "Goodman, Jessica and Ibe, Masahiro and Rajaraman, Arvind and Shepherd, William and Tait, Tim M. P. and Yu, Hai-Bo",
    title = "{Constraints on Dark Matter from Colliders}",
    eprint = "1008.1783",
    archivePrefix = "arXiv",
    primaryClass = "hep-ph",
    reportNumber = "UCI-HEP-TR-2010-15",
    doi = "10.1103/PhysRevD.82.116010",
    journal = "Phys. Rev. D",
    volume = "82",
    pages = "116010",
    year = "2010"
}

@article{Fermi-LAT:2016uux,
    author = "Albert, A. and others",
    collaboration = "Fermi-LAT, DES",
    title = "{Searching for Dark Matter Annihilation in Recently Discovered Milky Way Satellites with Fermi-LAT}",
    eprint = "1611.03184",
    archivePrefix = "arXiv",
    primaryClass = "astro-ph.HE",
    reportNumber = "FERMILAB-PUB-16-073-AE",
    doi = "10.3847/1538-4357/834/2/110",
    journal = "Astrophys. J.",
    volume = "834",
    number = "2",
    pages = "110",
    year = "2017"
}

@article{Bertone:2018krk,
    author = "Bertone, Gianfranco and Tait, M. P., Tim",
    title = "{A new era in the search for dark matter}",
    eprint = "1810.01668",
    archivePrefix = "arXiv",
    primaryClass = "astro-ph.CO",
    doi = "10.1038/s41586-018-0542-z",
    journal = "Nature",
    volume = "562",
    number = "7725",
    pages = "51--56",
    year = "2018"
}

@article{Nesti:2023tid,
    author = "Nesti, Fabrizio and Salucci, Paolo and Turini, Nicola",
    title = "{The Quest for the Nature of the Dark Matter: The Need of a New Paradigm}",
    eprint = "2308.02004",
    archivePrefix = "arXiv",
    primaryClass = "hep-ph",
    doi = "10.3390/astronomy2020007",
    journal = "Astronomy",
    volume = "2",
    number = "2",
    pages = "90--104",
    year = "2023"
}

@article{WMAP:2012nax,
    author = "Hinshaw, G. and others",
    collaboration = "WMAP",
    title = "{Nine-Year Wilkinson Microwave Anisotropy Probe (WMAP) Observations: Cosmological Parameter Results}",
    eprint = "1212.5226",
    archivePrefix = "arXiv",
    primaryClass = "astro-ph.CO",
    doi = "10.1088/0067-0049/208/2/19",
    journal = "Astrophys. J. Suppl.",
    volume = "208",
    pages = "19",
    year = "2013"
}

@article{Bergstrom:2000pn,
    author = {Bergstr\"om, Lars},
    title = "{Nonbaryonic dark matter: Observational evidence and detection methods}",
    eprint = "hep-ph/0002126",
    archivePrefix = "arXiv",
    doi = "10.1088/0034-4885/63/5/2r3",
    journal = "Rept. Prog. Phys.",
    volume = "63",
    pages = "793",
    year = "2000"
}

@article{PhysRevD.107.115016,
  title = {Bounds on long-lived dark matter mediators from neutron stars},
  author = {Nguyen, Thong T. Q. and Tait, Tim M. P.},
  journal = {Phys. Rev. D},
  volume = {107},
  issue = {11},
  pages = {115016},
  numpages = {8},
  year = {2023},
  month = {Jun},
  publisher = {American Physical Society},
  doi = {10.1103/PhysRevD.107.115016},
  url = {https://link.aps.org/doi/10.1103/PhysRevD.107.115016}
}

@ARTICLE{2024arXiv240502393A,
       author = {{Acevedo}, Javier F. and {Leane}, Rebecca K. and {Reilly}, Aidan J.},
        title = "{Dark Kinetic Heating of Exoplanets and Brown Dwarfs}",
      journal = {arXiv e-prints},
         year = 2024,
        month = may,
          eid = {arXiv:2405.02393},
        pages = {arXiv:2405.02393},
          doi = {10.48550/arXiv.2405.02393},
archivePrefix = {arXiv},
       eprint = {2405.02393},
 primaryClass = {astro-ph.EP},
       adsurl = {https://ui.adsabs.harvard.edu/abs/2024arXiv240502393A}
}

@article{Abbott:1982af,
    author = "Abbott, L. F. and Sikivie, P.",
    editor = "Srednicki, M. A.",
    title = "{A Cosmological Bound on the Invisible Axion}",
    reportNumber = "PRINT-82-0695 (BRANDEIS)",
    doi = "10.1016/0370-2693(83)90638-X",
    journal = "Phys. Lett. B",
    volume = "120",
    pages = "133--136",
    year = "1983"
}

@ARTICLE{1983PhLB..120..137D,
       author = {{Dine}, Michael and {Fischler}, Willy},
        title = "{The not-so-harmless axion}",
      journal = {Physics Letters B},
         year = 1983,
        month = jan,
       volume = {120},
       number = {1-3},
        pages = {137-141},
          doi = {10.1016/0370-2693(83)90639-1},
       adsurl = {https://ui.adsabs.harvard.edu/abs/1983PhLB..120..137D}
}

% \clearpage
% \onecolumngrid
\newpage
\appendix

\section{Exoplanets list}\label{exoplanets list}

\begin{table}[H]
    \centering
    \rotatebox{90}{
        \begin{minipage}{1\textwidth}
            \renewcommand{\arraystretch}{1.5}
            \setlength{\tabcolsep}{3pt}
            \begin{tabular}{lccccccc}
                \toprule
                Planet & Radius ($R_{\text{jup}}$) & Mass ($M_{\text{jup}}$) & $Age_{\text{host}}$(Gyr) & Orbit (au) & Temperature (K) & $M_{\text{host}}(M_{\odot})$ & $R_{\text{host}}(R_{\odot})$ \\
                \midrule
                Epsilon Eridani b \cite{Baines_2012}    & 1.21  & 1.55  & 0.8     & 5.2   & \( \lesssim 200 \)  & 0.82  & 0.738  \\
                Epsilon Indi A b \cite{feng2018detectionclosestjovianexoplanet} & 1.17  & 7.0  & 3.5   & 11.6  & \( \lesssim 200 \)  & 0.713  & 0.782  \\
                Gliese 832 b \cite{Bailey_2009} & 1.25  & 0.68  & 6.0   & 3.6   & \( \lesssim 200 \)  & 0.441  & 0.442  \\
                Gliese 849 b \cite{Butler_2006} & 1.23  & 1.0   & 3.0   & 2.4   & \( \lesssim 200 \)  & 0.465  & 0.464  \\
                Lipperhey \cite{Marcy_2002}     & 1.16  & 3.9   & 8.6   & 5.5   & \( \lesssim 200 \)  & 0.905  & 0.980  \\
                Gliese 777 b \cite{10.1111/j.1365-2966.2009.16233.x} & 1.21  & 1.54  & 4.79  & 4.0   & \( \lesssim 200 \)  & 1.142  & 0.93  \\
                Gliese 317 c \cite{Johnson_2007} & 1.21  & 1.54  & 5.0   & 25.0  & \( \lesssim 200 \)  & 0.42  & 0.417  \\
                HD 87883 b \cite{Fischer_2009}   & 1.21  & 1.54  & 7.6   & 3.6   & \( \lesssim 200 \)  & 0.80  & 0.76  \\
                HD 29021 b \cite{Rey2017}        & 1.2   & 2.4   & 7.4   & 2.3   & \( \lesssim 200 \)  & 0.85  & 0.85  \\
                Xolotan  \cite{Vogt_2002}        & 1.2   & 0.9   & 3.813 & 1.7   & \( \lesssim 200 \)  & 0.883  & 0.846  \\
                HAT-P-11c \cite{Yee_2018}        & 1.2   & 1.6   & 6.5   & 4.1   & \( \lesssim 200 \)  & 0.81  & 0.683  \\
                Pirx \cite{Niedzielski_2009}     & 1.2   & 1.1   & 6.9   & 0.8   & \( \sim 200 \)  & 0.82  & 0.78  \\
                HD 27631 b \cite{Marmier2013}    & 1.2   & 1.5   & 4.01  & 3.2   & \( \lesssim 200 \)  & 0.944  & 0.923  \\
                HD 6718 b  \cite{Naef2010}       & 1.2   & 1.7   & 6.0   & 3.6   & \( \lesssim 200 \)  & 0.98  & 1.01  \\
                \bottomrule
            \end{tabular}
        \end{minipage}
    }
    \caption{Exoplanet parameters, including planetary mass, radius, host star age, orbital radius, observed planetary temperature, and host star mass and radius.}
    \label{tab:exoplanets}
\end{table}

\section{Simulation of exoplanets}\label{simulation_result}

\begin{figure}[H]
    \centering
    % 第一张子图
    \subfloat[]{
        \includegraphics[scale=0.31]{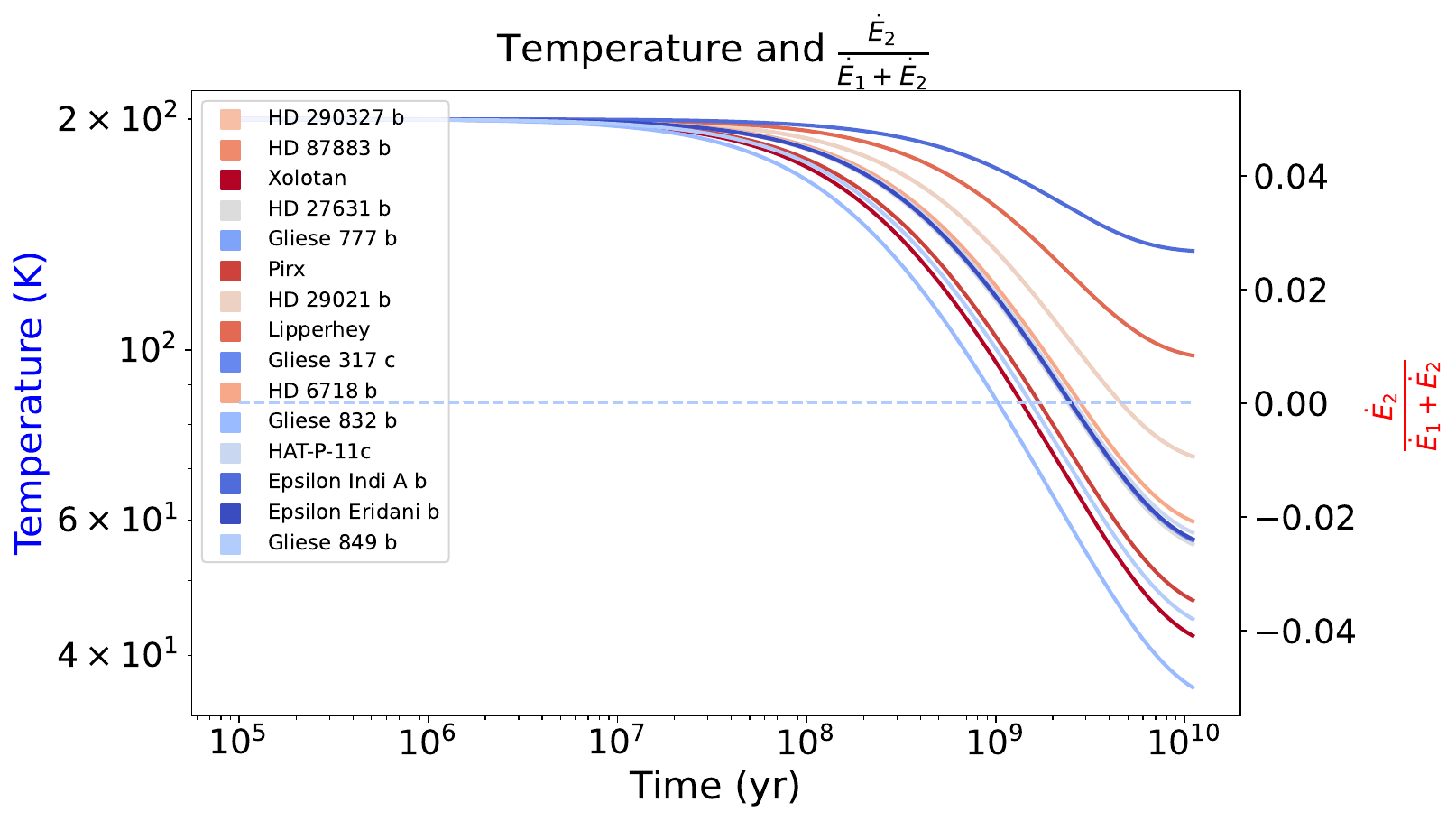}
        \label{a}

    }
    % 第二张子图
    \subfloat[]{

        \includegraphics[scale=0.31]{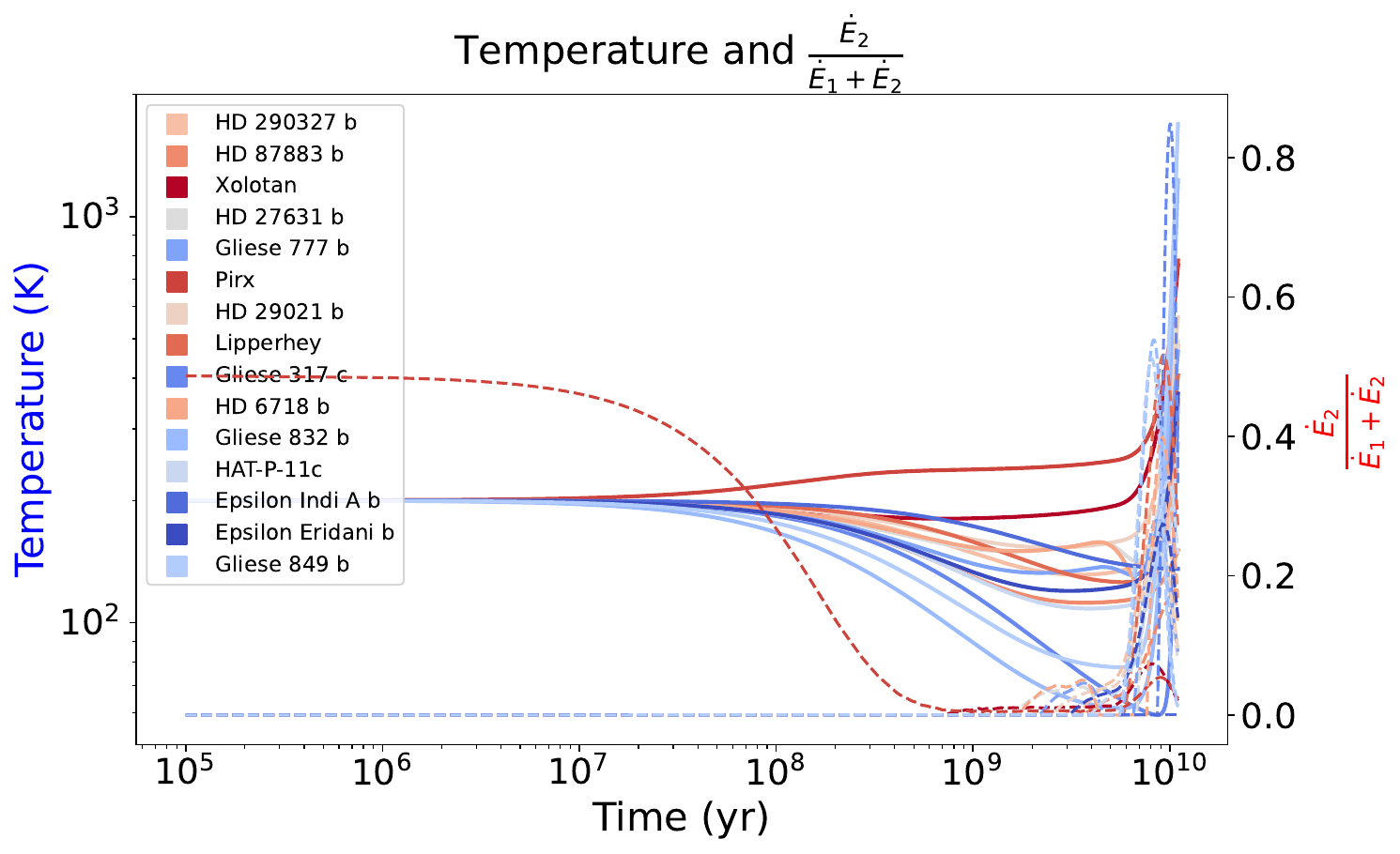}
    \label{b}
    }
    
    % 第3张子图
    \subfloat[]{
        \includegraphics[scale=0.31]{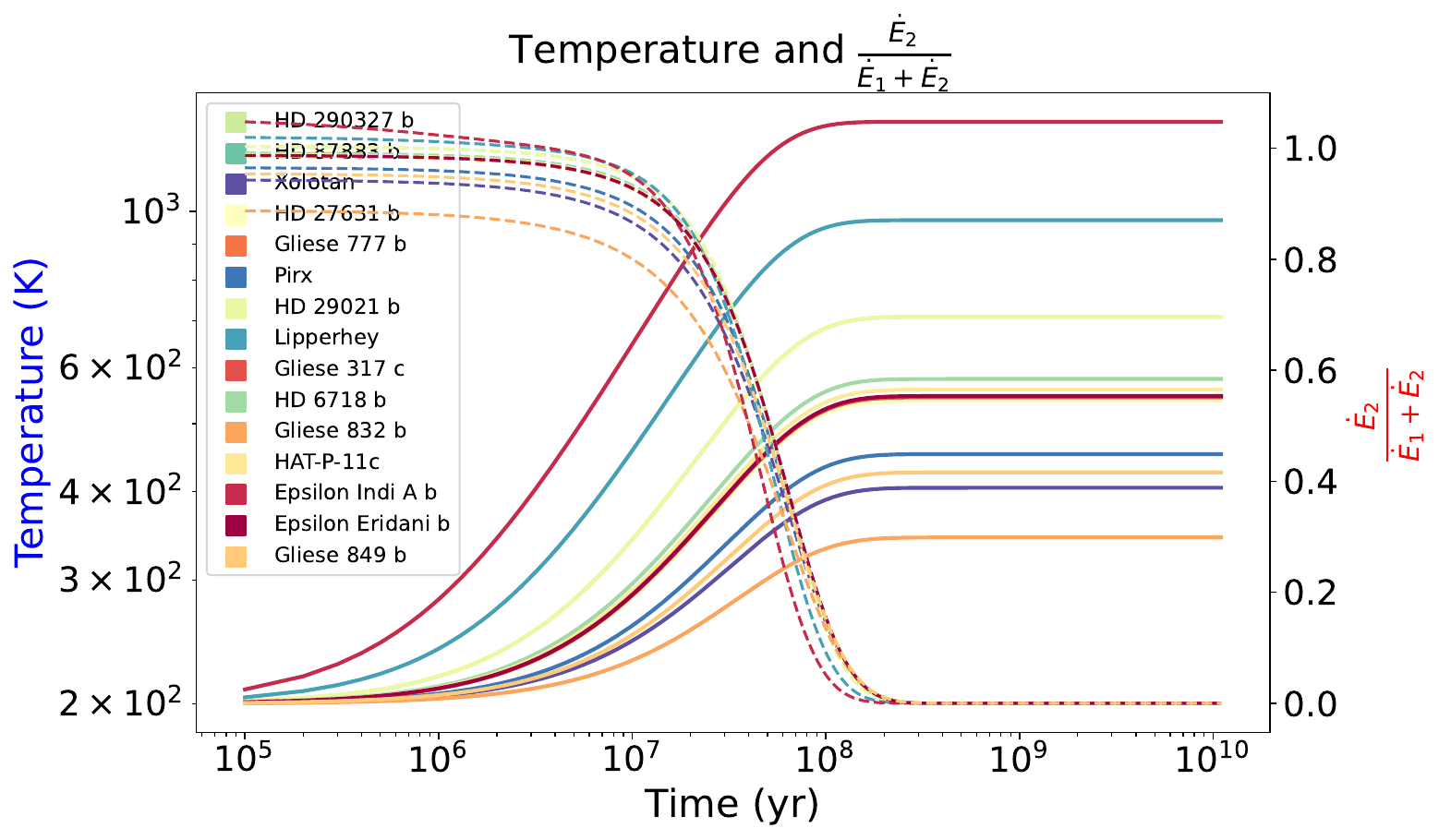}
        \label{c}
    }
    % 第4张子图
    \subfloat[]{

        \includegraphics[scale=0.31]{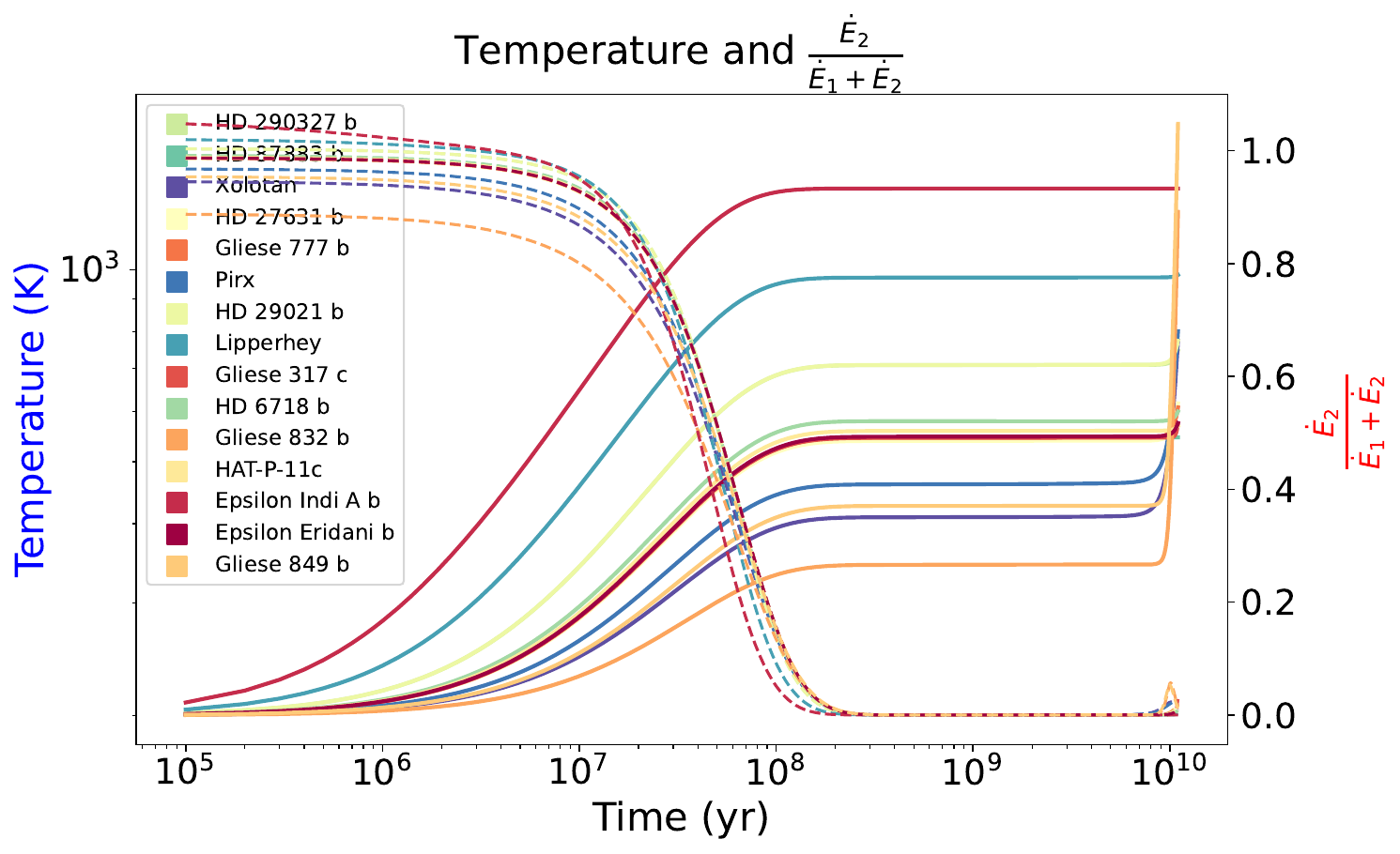}
    \label{d}
    }
    \caption{\autoref{a} shows the evolution of planetary temperature and acceleration energy ratio under the dark matter heating mechanism at a dark matter density of \(\rho_{\text{local}}\). \autoref{b} presents the evolution considering both solar and dark matter heating mechanisms at \(\rho_{\text{local}}\). \autoref{c} illustrates the evolution under the dark matter heating mechanism at a dark matter density of \(10^{4} \rho_{\text{local}}\). \autoref{d} depicts the evolution considering both solar and dark matter heating mechanisms at \(10^{4} \rho_{\text{local}}\).The solid line represents temperature, while the dashed line indicates the fraction of energy used to change angular velocity.}

    \label{2*2}
\end{figure}
\newpage
\section{Simulation of solar system}\label{solar_simu}

\begin{figure}[H]  
    \centering  
    % 第一张子图  
    \subfloat[]{  
        \includegraphics[scale=0.30]{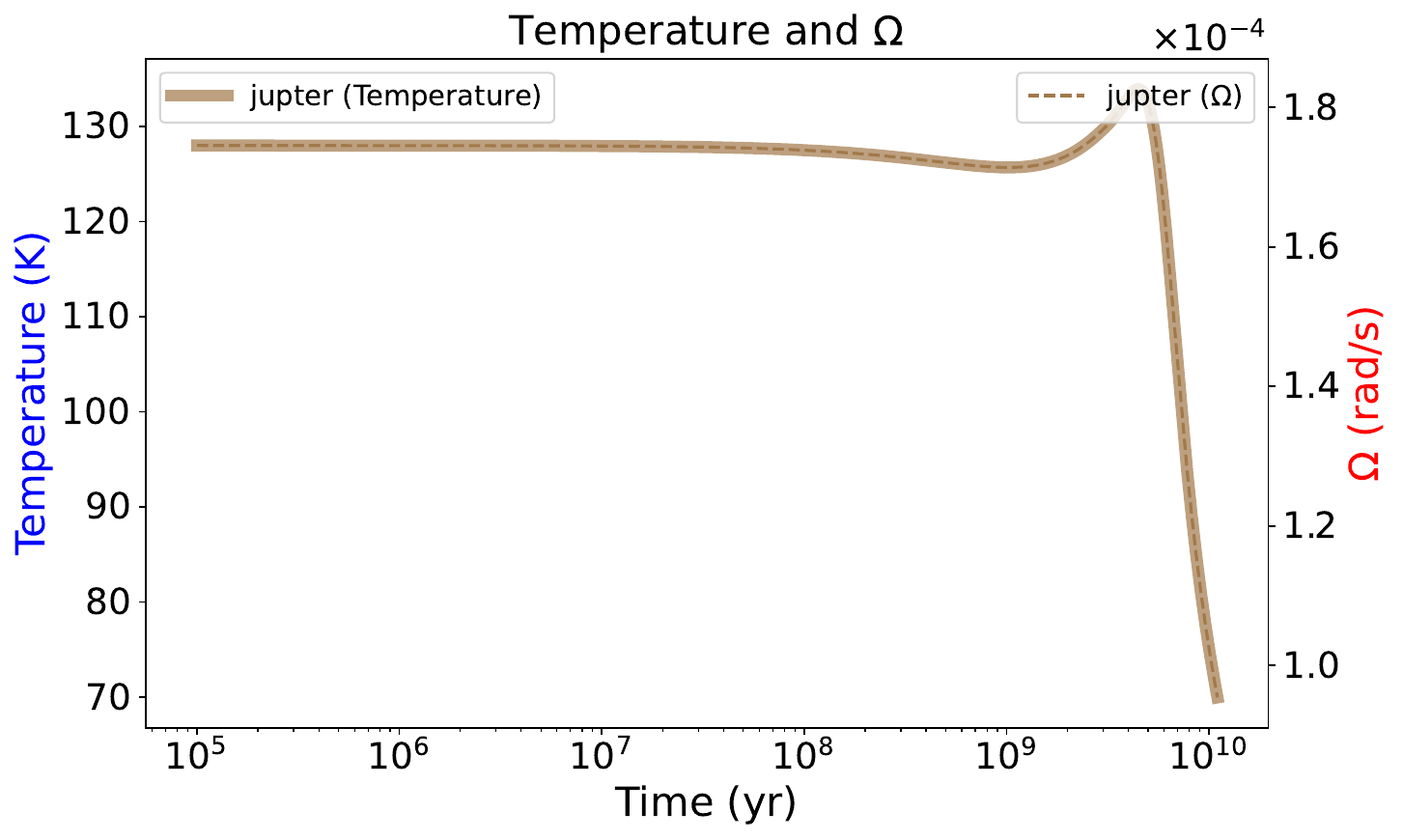}  
        \label{fig:jupiter_omega_long} }  
    % 第二张子图  
    \subfloat[]{  
        \includegraphics[scale=0.30]{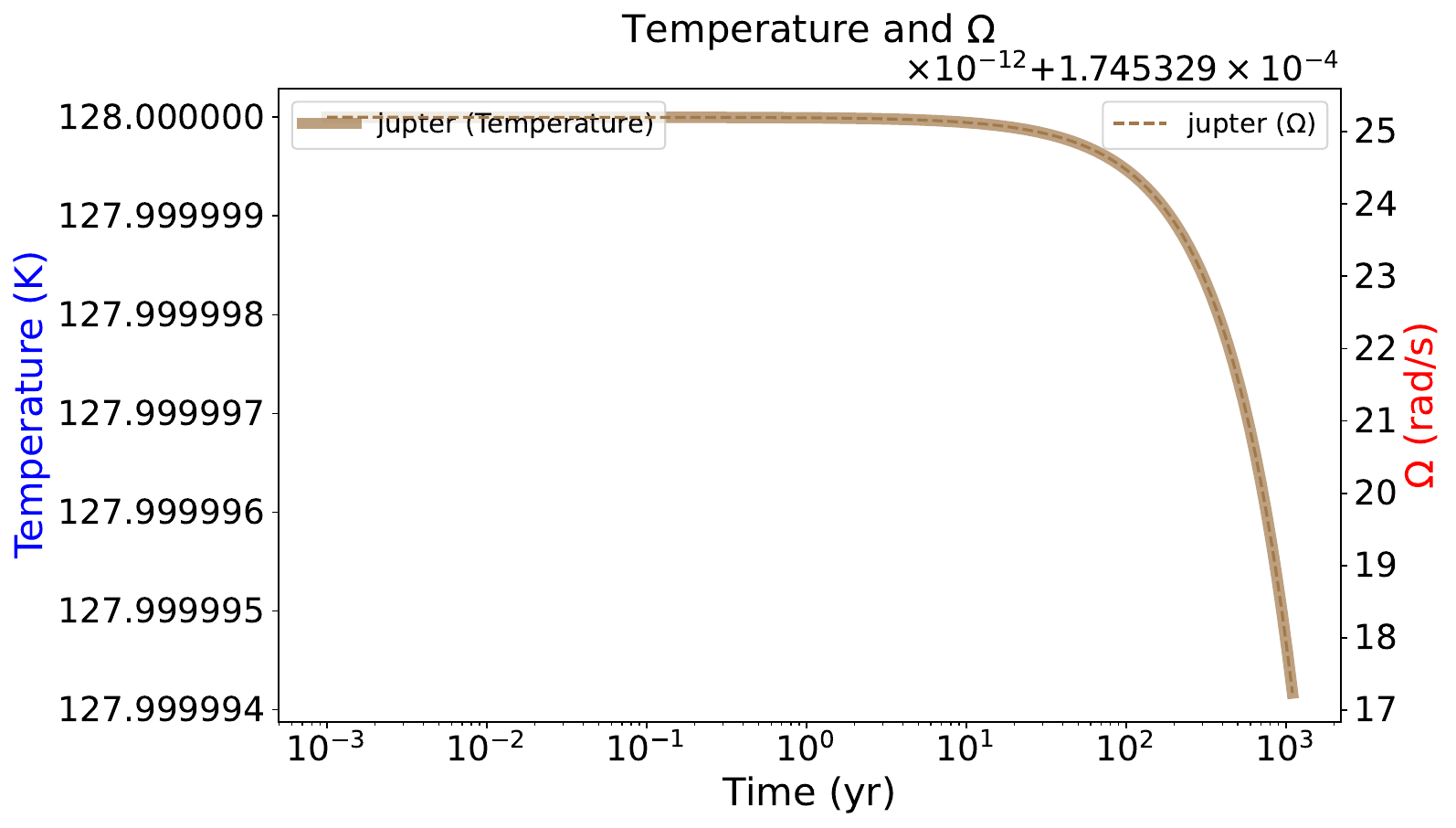}  
        \label{fig:jupiter_omega_short}  }  
    
    % 第三张子图  
    \subfloat[]{  
        \includegraphics[scale=0.30]{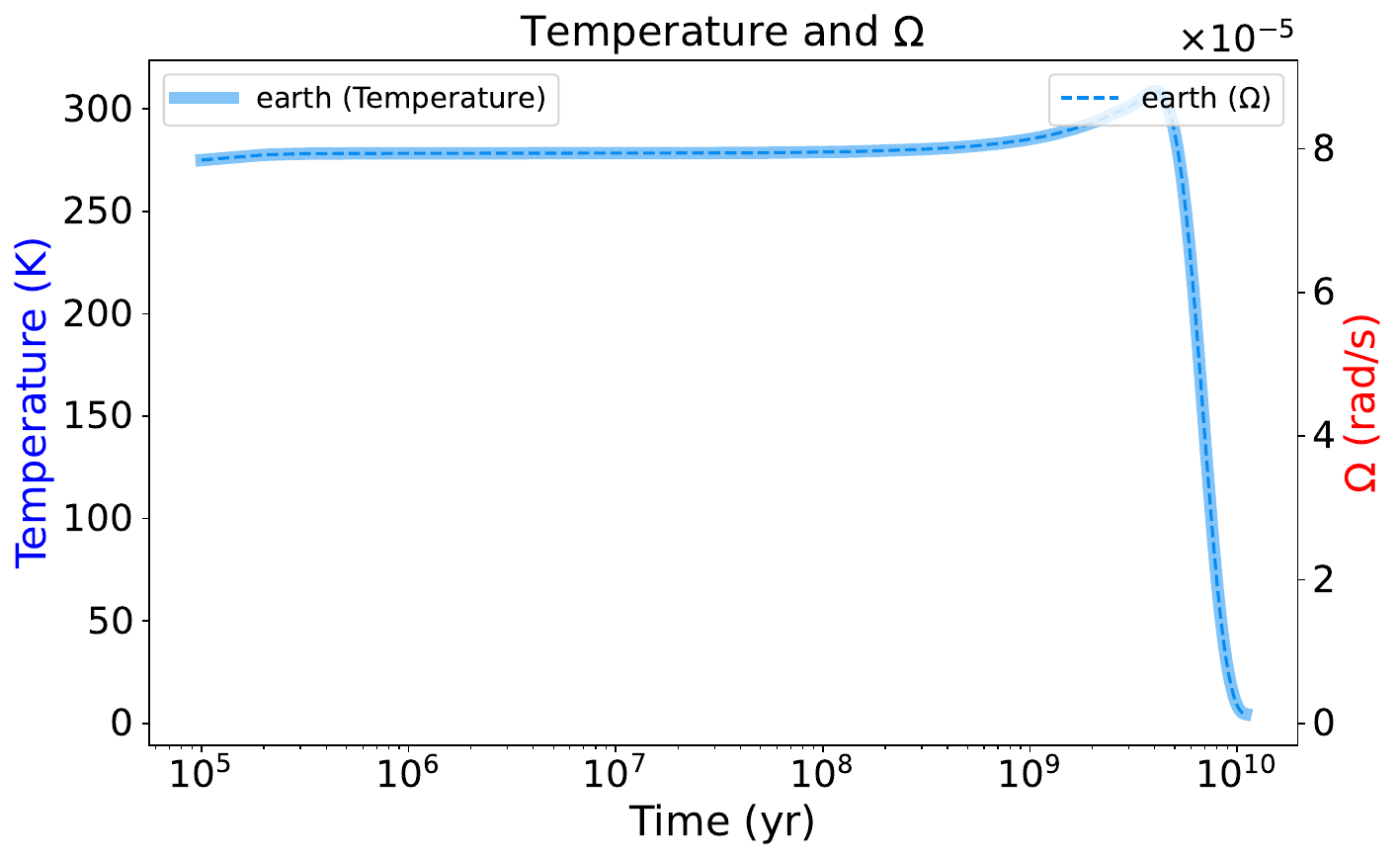}  
        \label{fig:earth_omega_long}  }  
    % 第四张子图  
    \subfloat[]{  
        \includegraphics[scale=0.30]{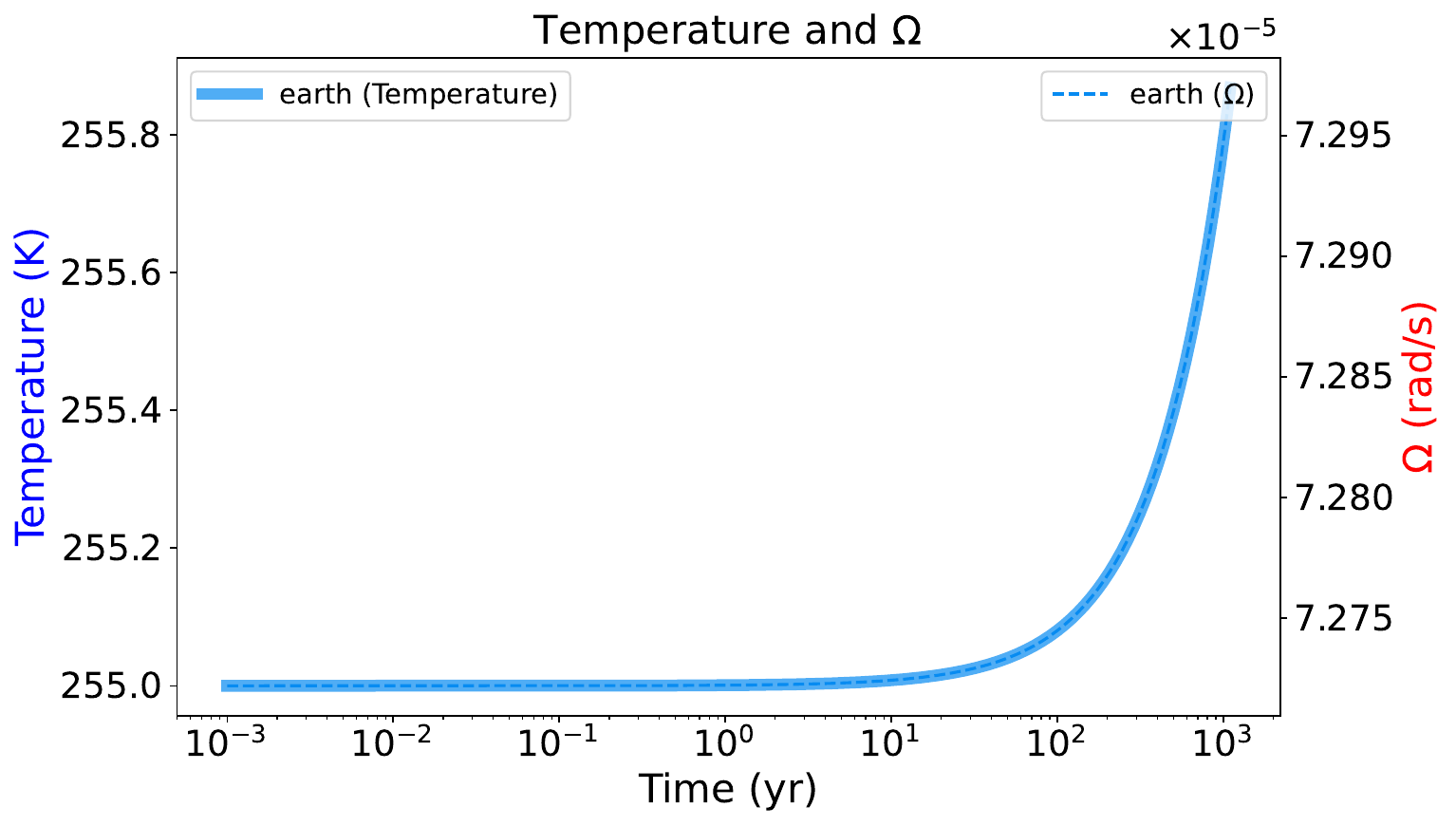}  
        \label{fig:earth_omega_short}  }  
    
    % 第五张子图  
    \subfloat[]{  
        \includegraphics[scale=0.30]{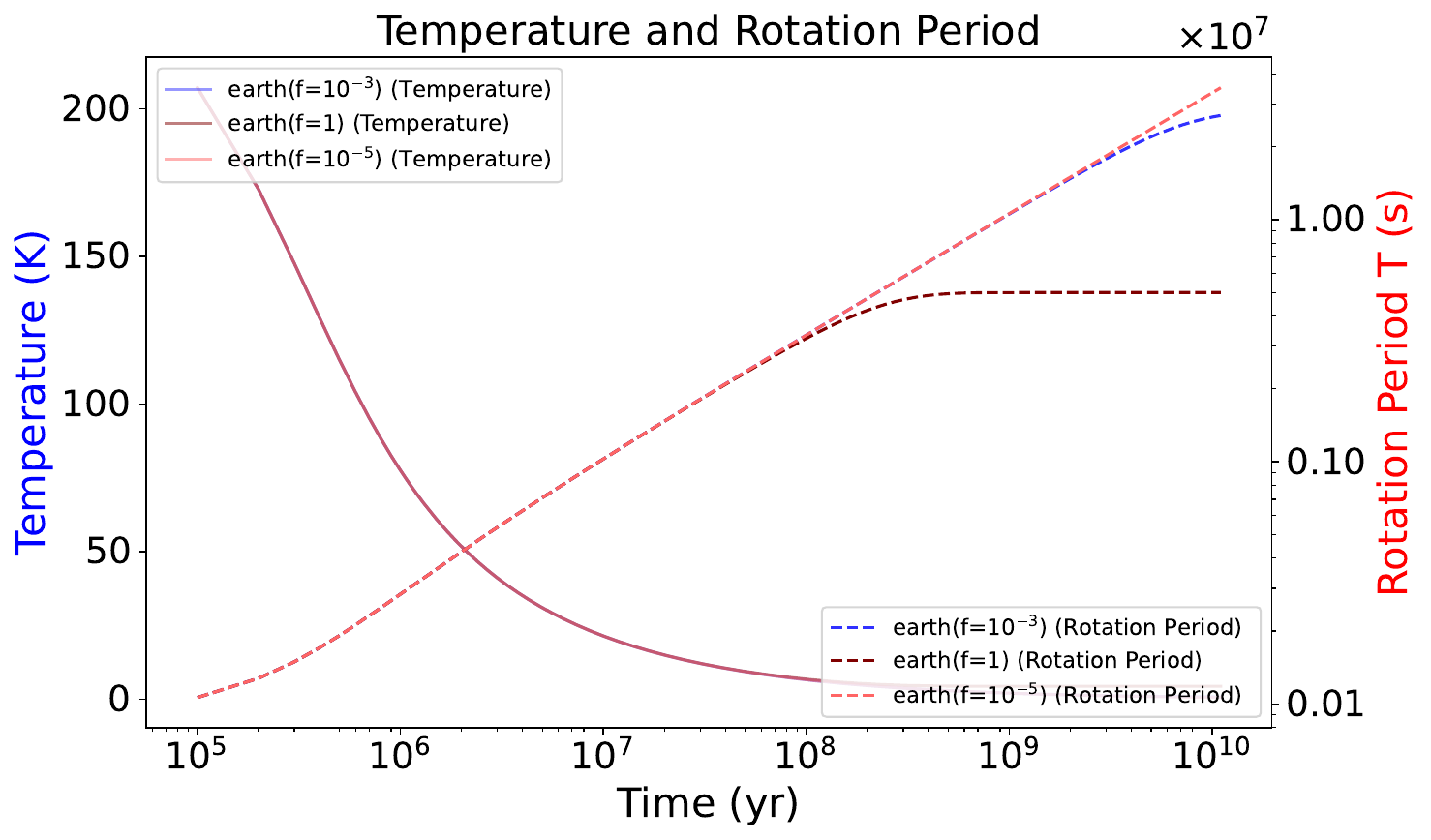}  
        \label{fig:earth_f_onlyDM}  
    }  
    % 第六张子图  
    \subfloat[]{  
        \includegraphics[scale=0.30]{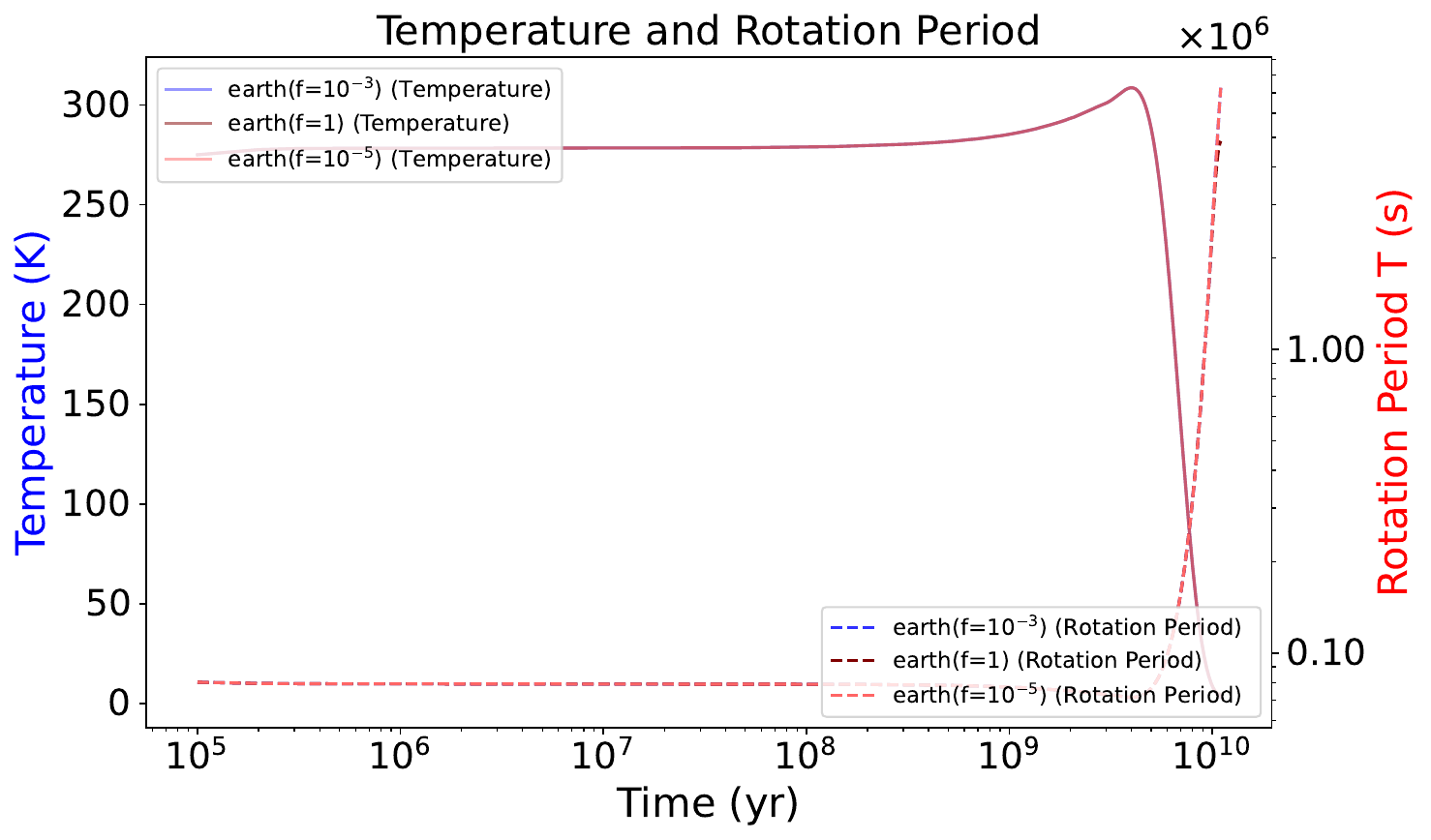}  
        \label{fig:earth_f_sun+DM}  }  
    
    \caption{\autoref{fig:jupiter_omega_long}, \autoref{fig:jupiter_omega_short}, \autoref{fig:earth_omega_long}, and \autoref{fig:earth_omega_short} illustrate the impact of our mechanism on the rotational dynamics of Jupiter and Earth over different timescales.
\autoref{fig:earth_f_onlyDM} shows the evolution of Earth's effective temperature and rotational angular velocity under dark matter heating alone, for $f$ values of $10^{-5}, 10^{-3}$, and 1 .
\autoref{fig:earth_f_sun+DM} presents the corresponding evolution with both dark matter and solar heating included, using the same $f$ values as in \autoref{fig:earth_f_onlyDM}.}

    \label{fig:evo}  
\end{figure}

% \acknowledgments

% This is the most common positions for acknowledgments. A macro is
% available to maintain the same layout and spelling of the heading.

% \paragraph{Note added.} This is also a good position for notes added
% after the paper has been written.

% Bibliography

%% [A] Recommended: using JHEP.bst file
%% \bibliographystyle{JHEP}
%% \bibliography{biblio.bib}

%% or
%% [B] Manual formatting (see below)
%% (i) We suggest to always provide author, title and journal data or doi:
%% in short all the informations that clearly identify a document.
%% (ii) please avoid comments such as "For a review'', "For some examples",
%% "and references therein" or move them in the text. In general, please leave only references in the bibliography and move all
%% accessory text in footnotes.
%% (iii) Also, please have only one work for each \bibitem.

% \begin{thebibliography}{99}

% \bibitem{a}
% Author,
% \emph{Title},
% \emph{J. Abbrev.} {\bf vol} (year) pg.

% \bibitem{b}
% Author,
% \emph{Title},
% arxiv:1234.5678.

% \bibitem{c}
% Author,
% \emph{Title},
% Publisher (year).

% \end{thebibliography}
\end{document}